%% file: nnetfix_paper.tex
\documentclass[12pt]{iopart}
\usepackage{cite}

\usepackage{iopams}  

\expandafter\let\csname equation*\endcsname\relax
\expandafter\let\csname endequation*\endcsname\relax
\usepackage{amsmath}
\usepackage{amssymb}
\usepackage{color}

\usepackage{graphicx}
\usepackage{lipsum} % for dummy text only
\usepackage{acronym,xspace}
\usepackage{macros}

\usepackage{array, multirow}   
\usepackage{booktabs}
\usepackage{tabularx}
\usepackage{rotating}
\usepackage{lscape} 
\usepackage{geometry}
\usepackage{color}

\usepackage{hyperref}
\hypersetup{
    colorlinks=true,
    linkcolor=blue,
    filecolor=magenta,      
    urlcolor=cyan,
}

\begin{document}

\title[NNETFIX]{NNETFIX: An artificial neural network-based denoising engine for gravitational-wave signals}

\input{author_list}

\vspace{10pt}
\begin{indented}
\item[]\today
\end{indented}

\input{abstract}

\maketitle

\input{introduction}
\input{methodology}

\input{evaluation_reconstruction_in_timeseries}
\input{effect_skylocalization}
\input{conclusion}
\input{acknowledgements}

\section*{References}
\bibliographystyle{iopart-num}
\bibliography{bibliography}

\input{evaluation_reconstruction_in_timeseries_tables.tex}
\input{effect_skylocalization_tables.tex}
\end{document}

%% file: author_list.tex
\author{Kentaro Mogushi}
%\ead{mogus261@gmail.com}

\address{Institute of Multi-messenger Astrophysics and Cosmology, Missouri University of Science and Technology, Physics Building, 1315 N.\ Pine St., Rolla, MO 65409, USA}

\author{Ryan Quitzow-James}
%\ead{ryan.quitzow-james@ligo.org}

\address{Institute of Multi-messenger Astrophysics and Cosmology, Missouri University of Science and Technology, Physics Building, 1315 N.\ Pine St., Rolla, MO 65409, USA}

\author{Marco Cavagli\`a}
%\ead{}

\address{Institute of Multi-messenger Astrophysics and Cosmology, Missouri University of Science and Technology, Physics Building, 1315 N.\ Pine St., Rolla, MO 65409, USA}

\author{Sumeet Kulkarni}
%\ead{sskulkar@go.olemiss.edu}

\address{Department of Physics and Astronomy, University of Mississippi, MS 38677-1848, USA}

\author{Fergus Hayes}
%\ead{}

\address{SUPA, School of Physics and Astronomy, University of Glasgow, Glasgow G12 8QQ, United Kingdom}

%% file: abstract.tex
\begin{abstract}
Instrumental and environmental transient noise bursts in gravitational-wave detectors, or glitches, may impair astrophysical observations by adversely affecting the sky localization and the parameter estimation of gravitational-wave signals. Denoising of detector data is especially relevant during low-latency operations because electromagnetic follow-up of candidate detections requires accurate, rapid sky localization and inference of astrophysical sources. NNETFIX is a machine learning-based algorithm designed to remove glitches detected in coincidence with transient gravitational-wave signals. NNETFIX uses artificial neural networks to estimate the portion of the data lost due to the presence of the glitch, which allows the recalculation of the sky localization of the astrophysical signal.
The sky localization of the denoised data may be significantly more accurate than the sky localization obtained from the original data or by removing the portion of the data impacted by the glitch.
We test NNETFIX in simulated scenarios of binary black hole coalescence signals and discuss the potential for its use in future low-latency LIGO-Virgo-KAGRA searches.
In the majority of cases for signals with a high signal-to-noise ratio, we find that the overlap of the sky maps obtained with the denoised data and the original data is better than the overlap of the sky maps obtained with the original data and the data with the glitch removed.
\end{abstract}

%% file: introduction.tex
\section{Introduction}\label{introduction}

The field of \ac{GW} astronomy began with the first direct detection of a \ac{GW} signal from a \ac{BBH} merger \cite{Abbott:2016GW150914} on September 14$^{\rm th}$, 2015. Nine additional
\ac{BBH} mergers were detected with high confidence during the first and second LIGO \cite{advLigo2015} and Virgo~\cite{advVirgo2015} observation runs \cite{GWTC-1}.
During the third LIGO-Virgo observation run,
39 binary merger events were detected with high confidence \cite{Abbott:2020niy},
including two exceptional \ac{BBH} events \cite{LIGOScientific:2020stg, PhysRevLett.125.101102, Abbott_2020} and a possible \ac{NSBH} merger \cite{Abbott:2020khf}.

On August 17$^{\rm th}$, 2017, the first detection of a \ac{GW} signal from a \ac{BNS} merger, GW170817, expanded multi-messenger astronomy to include \ac{GW} observations \cite{GW170817}. A short
\ac{GRB} was detected approximately 1.7 seconds after the \ac{BNS} merger time \cite{GBM:2017lvd}. The sky map calculated from the \ac{GW} signal allowed the identification of the event with
an \ac{EM} counterpart \cite{GW170817, GBM:2017lvd}. The association of this \ac{GW} event with the observed \ac{EM} transients supports the long-hypothesized model that at least some short
\acp{GRB} are due to \ac{BNS} coalescences \cite{Monitor:2017mdv} and has provided many insights into fundamental astrophysics and cosmology. In April 2020, a second \ac{BNS} merger without
an \ac{EM} counterpart was detected \cite{Abbott:2020uma}.

In order to detect \ac{GW} signals, ground-based \ac{GW} detectors must be extremely sensitive, causing them to become highly susceptible to instrumental and environmental noise
\cite{GWTC-1}. In particular, transient noise bursts, or {\em glitches}, may impair the quality of detector data. The presence of a glitch in the proximity of a \ac{GW} signal can adversely
affect the analysis of the latter, including calculating the sky localization of the source.
The most notable example of such an occurrence was GW170817, where the effect of a glitch was mitigated in low-latency by removing the contaminated portion of the data and in follow-up studies by applying ad hoc mitigation algorithms  \cite{GW170817,Driggers:2018gii}.

One possibility to mitigate the effect of a contaminating glitch would be to discard the data from the affected detector. This is the simplest and fastest solution; however, it is also likely to impact the analysis and sky localization, especially in cases where data is only available from two detectors.
Another technique that can be used in low-latency is {\em gating}, which removes the data affected by the glitch.
One method of gating is to set the data affected by the glitch to zero using a window function to smoothly transition into and out of the gate \cite{Usman:2015kfa}.
Gating was used in the case of GW170817 to produce the low-latency sky localization for \ac{EM} follow-up observations \cite{Pankow}. On larger latencies, glitch mitigation techniques such as using BayesWave \cite{Bayeswave} to model and subtract the glitch can be used \cite{Pankow}.
 
\begin{figure}[ht]
  \begin{center}
  \begin{minipage}[b]{0.49\textwidth}
    \includegraphics[width=\textwidth]{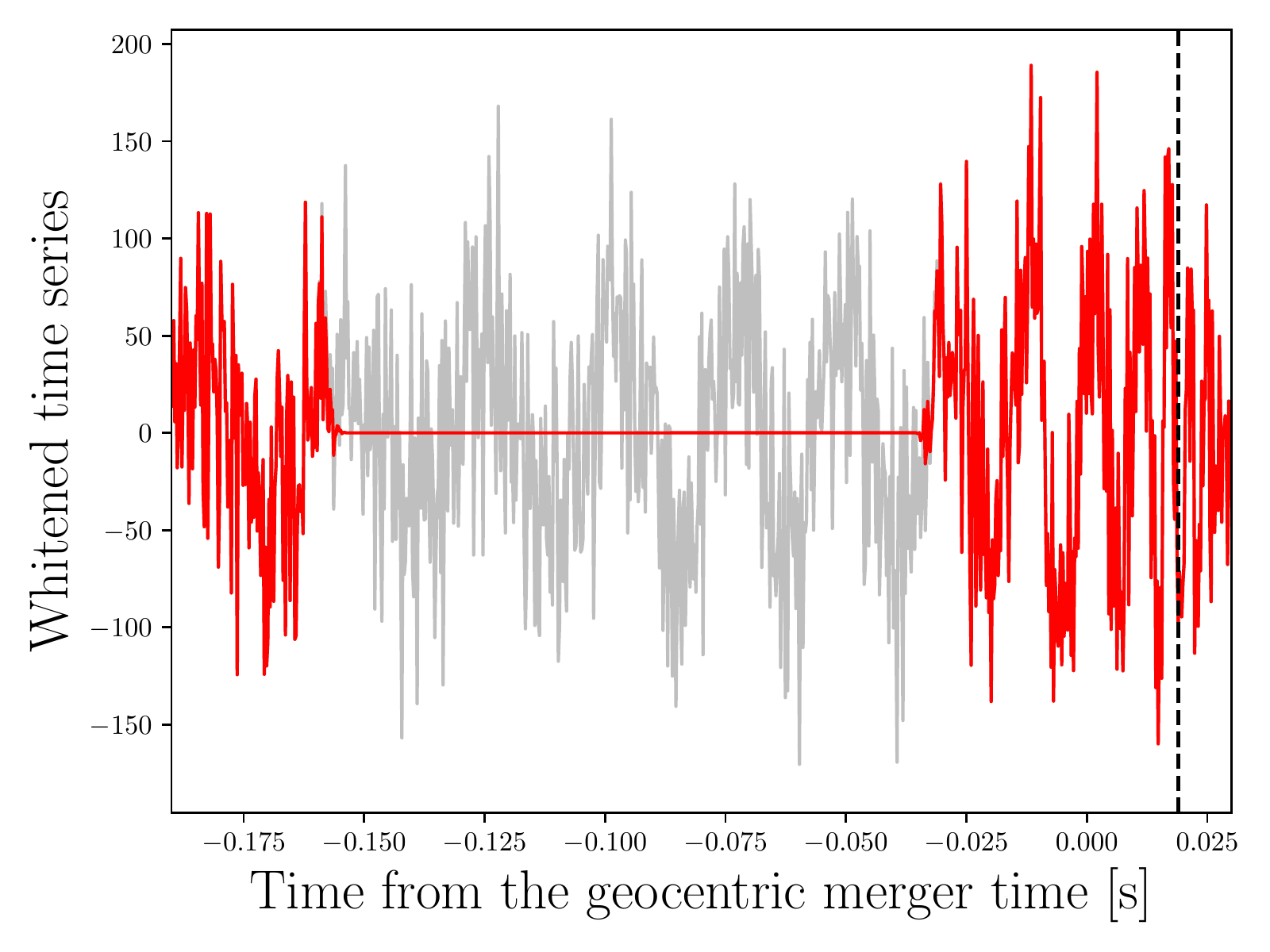}
  \end{minipage}
  \hfill
  \begin{minipage}[b]{0.49\textwidth}
    \includegraphics[width=\textwidth]{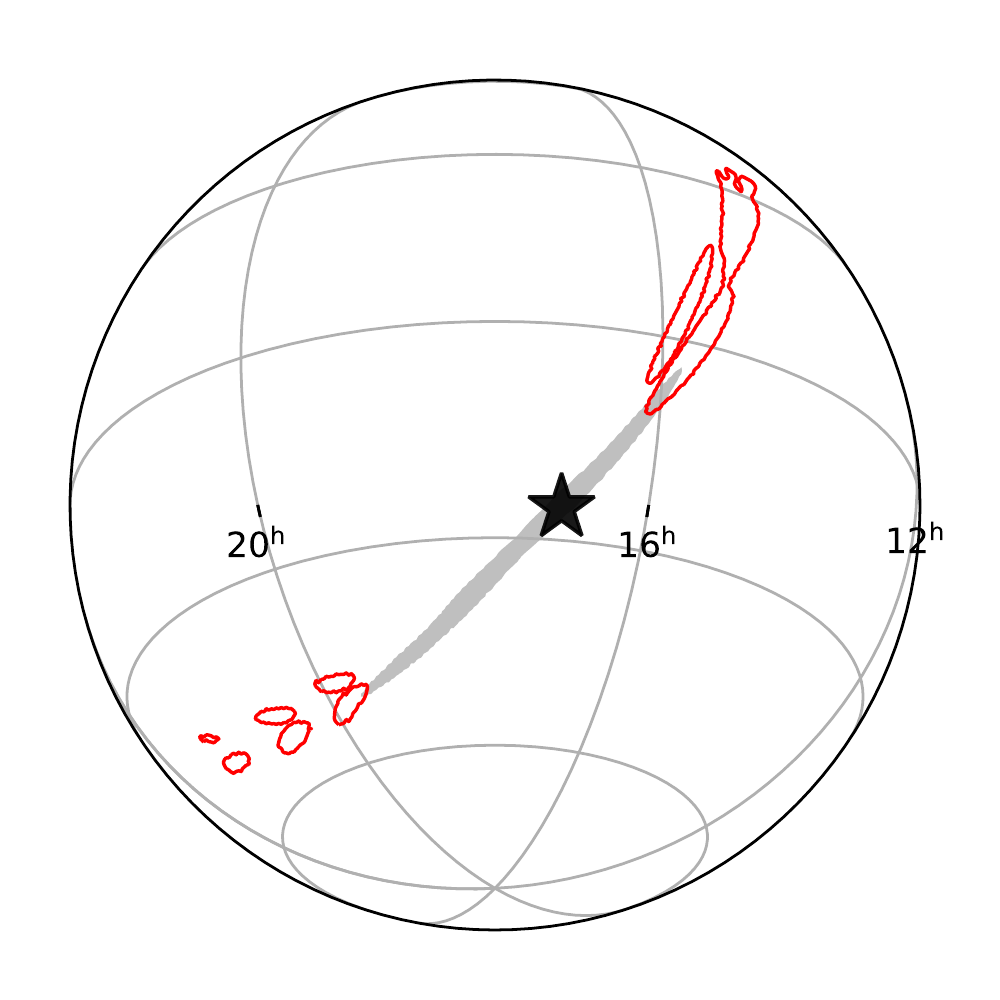}
  \end{minipage}
  \end{center}
  \caption{Left: Whitened time series of a simulated \ac{BBH} signal with two-detector network \ac{SNR} $\rho_{\rm N}=42.4$ and component masses $(m_1,m_2)=(35,29)$ \(M_\odot\) in 
    \ac{aLIGO} recolored Gaussian noise (red curve). A 130 ms gate is applied 30 ms before the geocentric merger time (gray curve). The vertical black-dashed line denotes the merger time in \ac{H1}. Right: The 90\% sky localization error regions 
    from the full data (gray area) and the gated data (red empty contours). The star indicates the true sky position of the simulated signal.}
 \label{fig:inj_result}
\end{figure}

Figure~\ref{fig:inj_result} shows an example of the detrimental effect gating data can have on the sky localization error region of a simulated \ac{BBH} merger signal. The sky localization obtained with the gated data
significantly differs from the sky localization estimated from the full data. Also, the 90\% sky localization error region after the gating is applied no longer includes the true sky position
of the injected signal.

The higher detector sensitivity of LIGO and Virgo in their third observing run has led to an increased number of \ac{GW} candidate detections from different astrophysical populations
\cite{LIGOScientific:2020stg, Abbott:2020uma, Abbott:2020khf}. Future observation runs with higher sensitivity are expected to produce even greater detection rates, which would lead to higher chances of observing \ac{GW} signals contaminated by glitches. The inability to accurately estimate the sky localization of \ac{GW} candidates with potential \ac{EM} counterparts due to glitches contaminating the signal could put at risk new
astrophysical discoveries such as those made with GW170817. Thus, the development and implementation of accurate low-latency denoising methods could be highly beneficial to multi-messenger
observations. 

In this paper, we present 
a machine learning-based algorithm to denoise transient \ac{GW} signals called
NNETFIX (``A Neural NETwork to `FIX' \ac{GW} signals coincident with short-duration glitches in detector data").
NNETFIX uses \acp{ANN} to estimate the portion of a signal which is lost due to
the presence of an overlapping glitch. We train the \ac{ANN} to reconstruct the portion of a gated signal on a template bank of \ac{BBH} waveforms injected into simulated noise data. The
accuracy of the algorithm is assessed by comparing the recovered waveform, \acf{SNR}, and sky map from the processed data to the corresponding quantities
obtained before gating. We derive a set of statistical metrics to assess the improvement in these quantities.

%% file: methodology.tex
\section{Algorithm implementation, training and testing} \label{methodology}

We consider a scenario in which a transient \ac{BBH} \ac{GW} signal is observed by a network of at least two detectors and the data of one detector is partially gated to remove a glitch. Without loss of generality, we perform the analysis for the two LIGO detectors, \acf{H1} and \ac{L1}, with the gating applied to data from the \ac{H1} detector. We assume the merger time at the geometric center of Earth (or geocentric merger time) to be (approximately) known from \ac{L1} data. We denote with $s_f(t)$, $s_g(t)$ and $s_r(t)$ the full time series, the gated time series, and the NNETFIX reconstructed time series, respectively. The output of NNETFIX can be thought of as the map
\begin{equation} \label{eq:predicted_ts}
    s_r(t) := F\left[s_g(t)\right]\,.
\end{equation}
We train an \ac{ANN} regression algorithm to construct the map $F$ such that $s_r(t)\sim s_f(t)$. The NNETFIX implementation uses the \textsc{scikit-learn} \cite{scikit-learn} \ac{MLP}
Regressor, a type of \ac{ANN} in which the nodes (mathematical functions) are arranged into layers and connected to every node in the preceding and/or succeeding layers \cite{MLP}. Each
node calculates a weighted linear combination of the outputs from the preceding layer and applies an activation function that introduces a non-linearity into the node's output.
The \ac{ANN} trained by NNETFIX consists of one hidden layer containing 200 neurons.
In the \ac{ANN} training process, NNETFIX uses the rectified linear unit (\textit{ReLU}) activation function \cite{Nair2010RectifiedLU} and the \textit{ADAM} stochastic gradient-based optimizer \cite{adam} with a learning rate of $10^{-3}$.
Ten percent of the training data samples are set aside and used for validating the training. The training iteration stops if the \ac{ANN} performance plateaus with a tolerance level of $10^{-4}$ to avoid overtraining.
To reconstruct the gated portion of the time series, one hidden layer works better than multiple hidden layers for the loss function of mean square error and the number of hidden layers tested.
The values from the loss function have a weak dependency on the number of neurons.

To train the algorithm, we first build template banks of simulated non-spinning \textsc{IMRphenomD} \ac{BBH} merger waveforms \cite{IMRPhenomD} with varying intrinsic and extrinsic
parameters. To reduce the potential for overtraining, each template bank also includes a number of (pure) noise time series. We distribute the positions of the injected
signals isotropically in the sky.
The waveform coalescence phase, polarization angle, and cosine of the inclination angle are uniformly distributed in the intervals $[0, 2\pi]$, $[0, \pi]$, and $[-1,1]$,
respectively. We uniformly distribute the network \ac{SNR} $\rho_{\rm N}$ \cite{Usman:2015kfa} of the simulated signals in the range $[11.3,42.4]$. We consider three distinct template banks
corresponding to low, medium, and high \ac{BBH} component masses to assess the prediction accuracy of the trained \acp{ANN} for different signal lengths. The \ac{BBH} component masses are uniformly sampled according to a Jeffreys prior for the matched-filter detection statistic. As
the mass of the system decreases, we employ a higher number of templates to properly cover the mass parameter space  \cite{Cokelaer:2007kx,Harry:2009ea,VanDenBroeck:2009gd,Manca:2009xw}. 

For each of the three distinct template banks, we build 12 \ac{TT} sets: first, we inject each waveform into 50 distinct realizations of recolored Gaussian noise for \acf{aLIGO} at design
sensitivity; second, we include the (pure) noise time series; third, we shuffle and split the set by 70\%--30\% for training and testing; and finally, we apply the 12 combinations of gate
durations $t_d=$ (50, 75, 130) ms and gate end-times before the geocentric merger time $t_e=$ (15, 30, 90, 170) ms. The time series are sampled at 4096 Hz, whitened, and high-passed.
A conservative value of 25 Hz is used for the high-pass filter. The gates are implemented as a reversed
Tukey window with a taper of 0.1 s and held fixed with respect to geocentric merger time; however, the merger time seen in the \ac{H1} detector naturally shifts due to the sky position and the
polarization angle of the \ac{GW} signal. Table \ref{mass_table} shows the range of the component masses, the number of waveforms, the number of noise series, and the dimension of the sets
for the different scenarios.

\begin{table}[ht]
\centering
\begin{tabular}[t]{l>{\centering}p{0.15\linewidth}>{\centering}p{0.15\linewidth}>{\centering}p{0.08\linewidth}>{\centering}p{0.08\linewidth}>{\centering\arraybackslash}p{0.3\linewidth}}
\toprule
     & $m_1$ [\(M_\odot\)] & $m_2$ [$M_\odot$] & $n_s$ & $n_n$ & Set dimension ($n_s\times 50 + n_n$)\\
\midrule
    Low   & 10--15 & 8--12 & 348 & 1900 & 19300\\
    Medium  & 15--25 & 12--18 & 251 & 1350 & 13900\\
    High  & 28--42 & 23--35 & 61 & 300 & 3350\\
\bottomrule
\end{tabular}
  \caption{Component mass ranges, number of waveforms ($n_s$), number of pure noise series ($n_n$), and dimension of the \ac{TT} sets for each of the three scenarios and combinations of gate durations and end-times.}\label{mass_table}
\end{table}

We test the effectiveness of the \acp{ANN} by calculating the coefficient of determination for the \ac{MLP} Regressor in \textsc{scikit-learn} on the testing sets \cite{scikit-learn}. The
coefficient of determination ranges from $-\infty$ (bad) to 1 (perfect estimation), with positive values corresponding to some degree of accuracy. We evaluate the coefficient of
determination on each testing set after training the \ac{ANN} on the corresponding training set. The ranges of the coefficient of determination for the testing sets are [0.773, 0.882],
[0.750, 0.883], and [0.691, 0.879] for the low-mass, medium-mass, and high-mass scenarios, respectively, and the means are 0.833, 0.827, and 0.814.

We test for potential statistical effects in the training method by considering the medium mass scenario with a gate duration of 50 ms and a gate end-time of 30 ms as a representative case.
For 100 trials, we find
that the coefficient of determination ranges from 0.800 to 0.826 with a mean of 0.815, which is consistent with the ranges of the testing sets.

The effect of NNETFIX on quantities such as \ac{SNR} and sky localization varies for different component masses, network \ac{SNR}, and gate settings. Therefore, we construct 108 additional independent {\em exploration} sets
with fixed network \ac{SNR} $\rho_{\rm N}=(11.3$, 28.3, $42.4)$ and component masses of (12, 10), (20, 15), (35, 29) $M_\odot$,
and identical combinations of gate durations and end-times as the \ac{TT} sets.
Each exploration set consists of 512 independent time series with the remaining parameters distributed as in the \ac{TT} sets.

%% file: evaluation_reconstruction_in_timeseries.tex
\section{Performance in the time-domain}\label{evaluation_in_time_series}

NNETFIX's performance in estimating the full time series can be assessed by computing the amount of \ac{SNR} lost in the reconstruction process. We define the
\ac{FRS} 
\begin{equation} \label{eq:snr_loss}
    \text{FRS} = \frac{\rho_f - \rho_r}{\rho_f}\,,
\end{equation}
where $\rho_f$ and $\rho_r$ are the (single interferometer) peak \ac{SNR} of the full time series and the reconstructed time series in \ac{H1}, respectively.
Positive values of \ac{FRS} close to zero generally indicate accurate time series reconstructions.
However, \ac{FRS} $\sim 0$ may also occur when the gating does not significantly reduce the peak \ac{SNR} of the full series,
and thus, $\rho_f\sim\rho_g\sim\rho_r$. These cases can be separated by the \ac{FSG}
\begin{equation} \label{eq:snr_gain}
    \text{FSG} = \frac{\rho_r - \rho_g}{\rho_g}\,,
\end{equation}
where $\rho_g$ is the peak \ac{SNR} of the gated series.
The \ac{FSG} characterizes the amount of \ac{SNR} gained by the reconstructed time series in comparison to the gated time series. Typically, NNETFIX performance is better for smaller values of \ac{FRS} and larger values of \ac{FSG}.

Median values of FRS across the exploration sets range from \ac{FRS} $=-0.09$ (high-mass case with $\rho_N=11.3$, $t_d=170$ ms and $t_e=50$ ms) to \ac{FRS} $=0.22$ (medium-mass case with
$\rho_N=28.3$, $t_d=15$ ms and $t_e=130$ ms). Sets with smaller gate durations are generally characterized by lower \ac{FRS} values. All exploration sets with $t_d=50$ ms have \ac{FRS}
$<0.1$. The fraction of sets with \ac{FRS} below this threshold reduces to 0.78 and 0.55 for gate durations of 75 ms and 130 ms, respectively. Similarly, exploration sets with gates that
are farther away from the time of the merger also tend to have a lower median value of the \ac{FRS}.
All sets with gate end-times at 170 ms before merger have median \ac{FRS} $<0.07$ while
only 81\%, 55\% and 11\% of the samples with $t_e=90$, 30 and 15 ms have \ac{FRS} below this threshold, respectively.
The effects of the network \ac{SNR} and component masses on \ac{FRS} are less significant. 
About 83\% of the sets with $\rho_N=11.3$ have median \ac{FRS} $<0.1$ compared to 75\% for the sets with $\rho_N=28.3$ and $\rho_N=42.4$.
The percentages of low-mass, medium-mass and high-mass sets with \ac{FRS} $<0.1$ are comparable at about 78\%, 75\% and 81\%, respectively.

Median values of \ac{FSG} range from $\sim 0.02$ (low-mass, low-\ac{SNR} case with $t_d=170$ ms and $t_e=50$ ms) to $\sim 0.89$ (high-mass, low-\ac{SNR} case with $t_d=130$ ms and $t_e=15$
ms). High-mass (low-mass) exploration sets are typically characterized by higher (lower) values of the \ac{FSG}. All high-mass sets have \ac{FSG} $>0.08$, while all low-mass sets have \ac{FSG} $<0.07$.
Gate end-time and \ac{SNR} do not seem to significantly affect the value of \ac{FSG}. The gate duration has a larger effect. Sets with longer gate durations typically have larger median
value of the \ac{FSG}. Two thirds of the sets with $t_d=130$ ms have \ac{FSG} $>0.08$ compared to only 42\% of the sets with $t_d=50$ ms. 

\begin{figure}[ht!]
    \centering
    \includegraphics[width=0.7\linewidth]{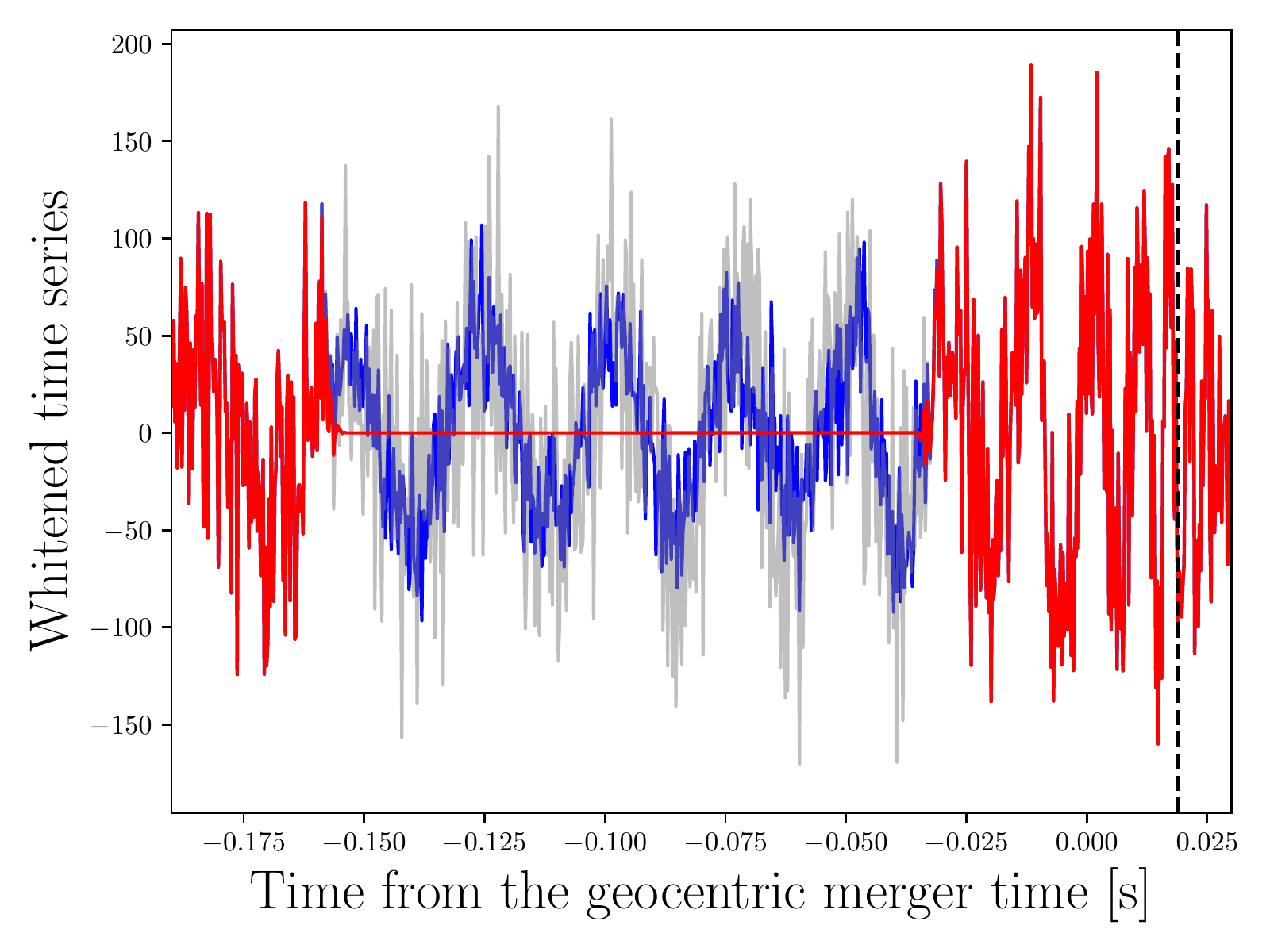}
    \caption{The whitened full time series (gray), the gated time series (red) and the reconstructed time series (blue) for the simulated event of Fig.~\ref{fig:inj_result}. The vertical black-dashed line denotes the merger time in \ac{H1}.}
    \label{fig:estimated_ts}
\end{figure}

\begin{figure}[ht!]
    \centering
    \includegraphics[width=\linewidth]{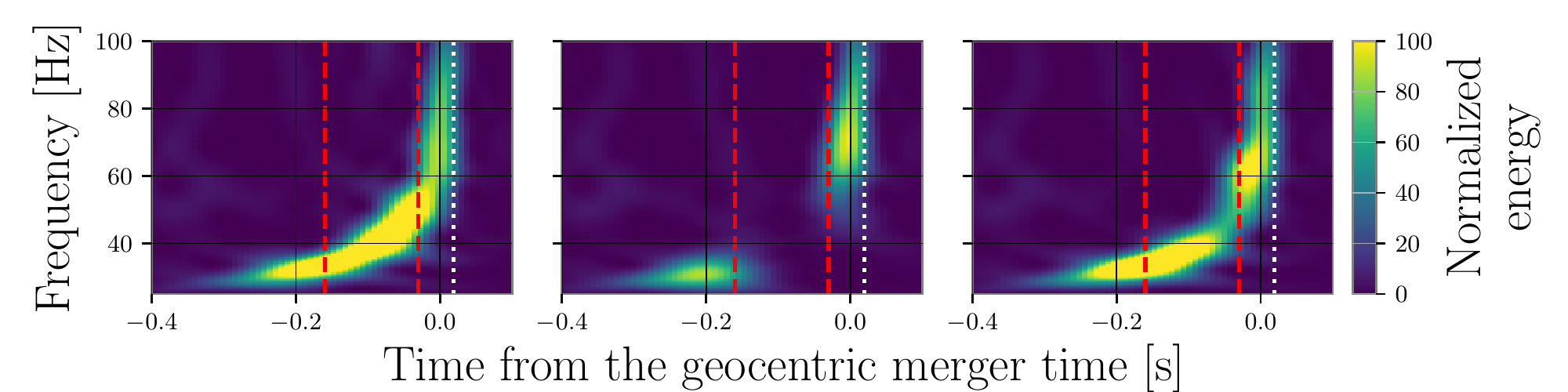}
    \caption{Time frequency representations of the full (left), gated (middle), and reconstructed (right) time series for the simulated event of Fig.~\ref{fig:estimated_ts} using the Q transform \cite{Chatterji:2004qg}. The vertical red-dashed line denotes the gate and the vertical white-dotted line denotes the merger time in \ac{H1}. }
    \label{fig:qtrasn}
\end{figure}

A combined threshold of \ac{FRS} $\gtrsim 0$ and \ac{FSG} $\gtrsim 0.01$ is a conservative choice for good reconstructions. About 70\% of the samples across all the exploration sets satisfy
this criterion. Figure~\ref{fig:estimated_ts} shows the NNETFIX data reconstruction for the time series of Fig.~\ref{fig:inj_result}. As shown in Fig.~\ref{fig:qtrasn}, the reconstructed time series recovers a large signal energy in the gated portion. In this case, \ac{FRS} $=0.05$ and \ac{FSG} $=0.53$.

A complementary metric to evaluate the performance of the algorithm is the \ac{FMG}
\begin{equation} \label{eq:gain_over_loss}
    \text{\ac{FMG}} = \frac{M_r - M_g}{M_f - M_g}\,,
\end{equation}
where the match $M_i$ between a time series $s_i$ and the injected waveform $h$ is
\begin{equation} \label{eq:match}
    M_i = \frac{\langle s_i|h\rangle}{\sqrt{\langle s_i|s_i\rangle\langle h|h\rangle}}\,.
\end{equation}
The inner product of two time series $s_i$ and $s_j$ is defined as
\begin{equation}
    \langle s_i|s_j\rangle = 4\Re  \int^{f_N}_{f_1}\frac{\tilde{s_i}(f)\tilde{s_j}^\ast(f)}{S(f)}df\,,
\end{equation}
where the tilde indicates the Fourier transform, the star denotes the complex conjugate, $S(f)$ is the detector noise \ac{PSD}, $f_1$ is the high-pass frequency, and $f_N$ is the Nyquist
frequency.

In Eq.~(\ref{eq:gain_over_loss}), we assume $M_f - M_g>0$. In rare instances (0.5\% of all exploration set data samples), $M_g$ becomes larger than $M_f$. This occurs for small values of the
single interferometer peak \ac{SNR} (median value of 4.6) when the gated portion of the data is dominated by noise and anti-correlates with the injected waveform. In the following, we
remove these data samples from the exploration sets.

\begin{figure}[ht!]
    \centering
    \includegraphics[width=0.7\linewidth]{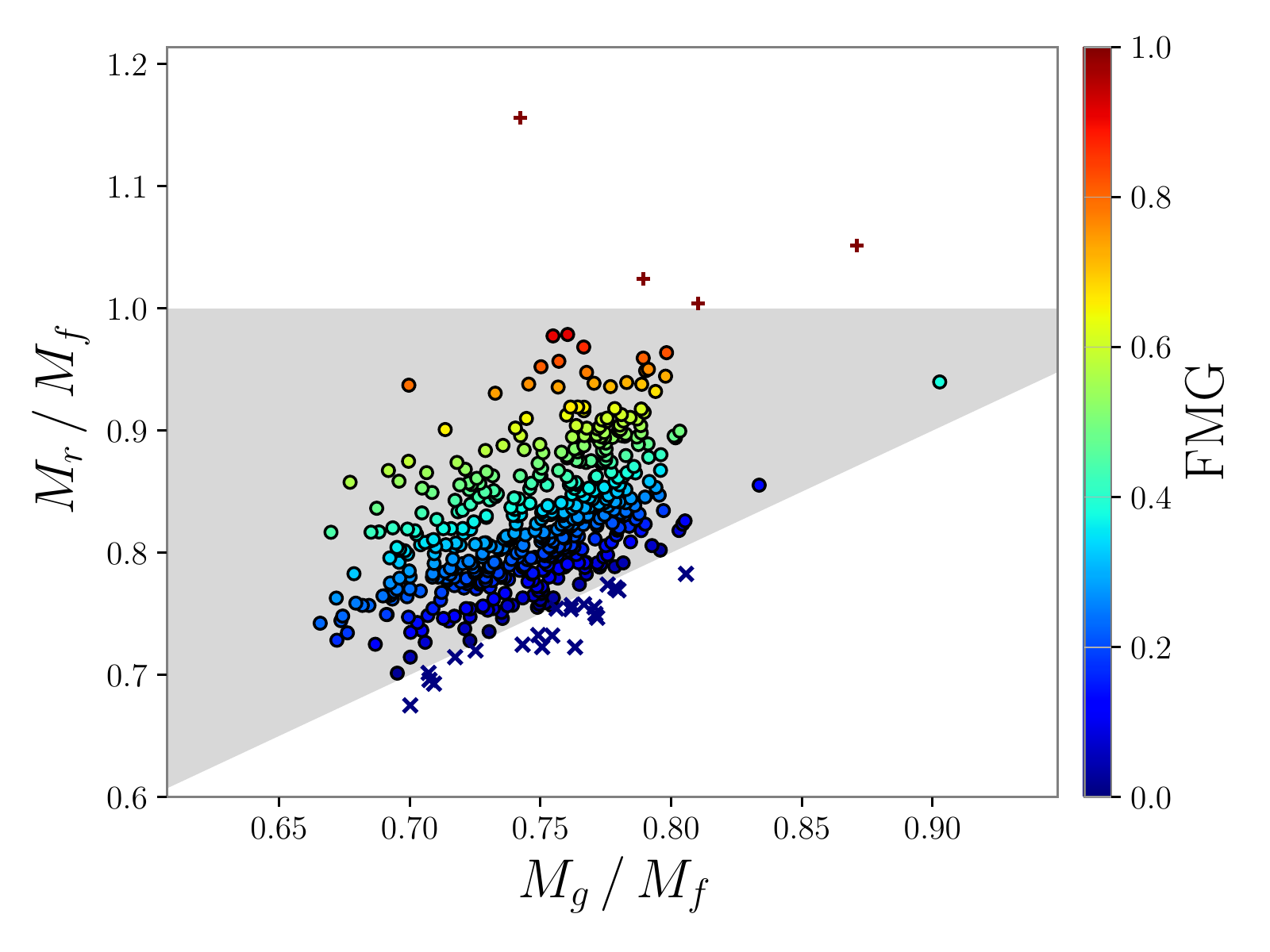}
    \caption{Scatterplot of $M_r/M_f$ vs.\ $M_g/M_f$ for the exploration set with $\rho_N=42.4$, $(m_1,m_2)=(20,15)$ \(M_\odot\), $t_d = 130$ ms and $t_e = 30$ ms. The circles denote samples with $0<\text{\ac{FMG}}\le 1$, the $\times$ markers denote samples with \ac{FMG} $\le 0$ and the $+$ markers denote overfitted samples with \ac{FMG} $>1$. The gray area denotes the region of the parameter space with $0<\text{\ac{FMG}}\le 1$, which contains 95\% of the reconstructed time series.
Two outliers with values \ac{FMG}$=-0.4$ and \ac{FMG}$=1.6$ are not shown in the plot in order to improve readability.}
    \label{fig:snr30_m1_20_m2_15_tbm_30ms_dur_130ms_roc_O_r_O_g}
\end{figure}

\begin{figure}[ht!]
    \centering
    \includegraphics[width=0.7\linewidth]{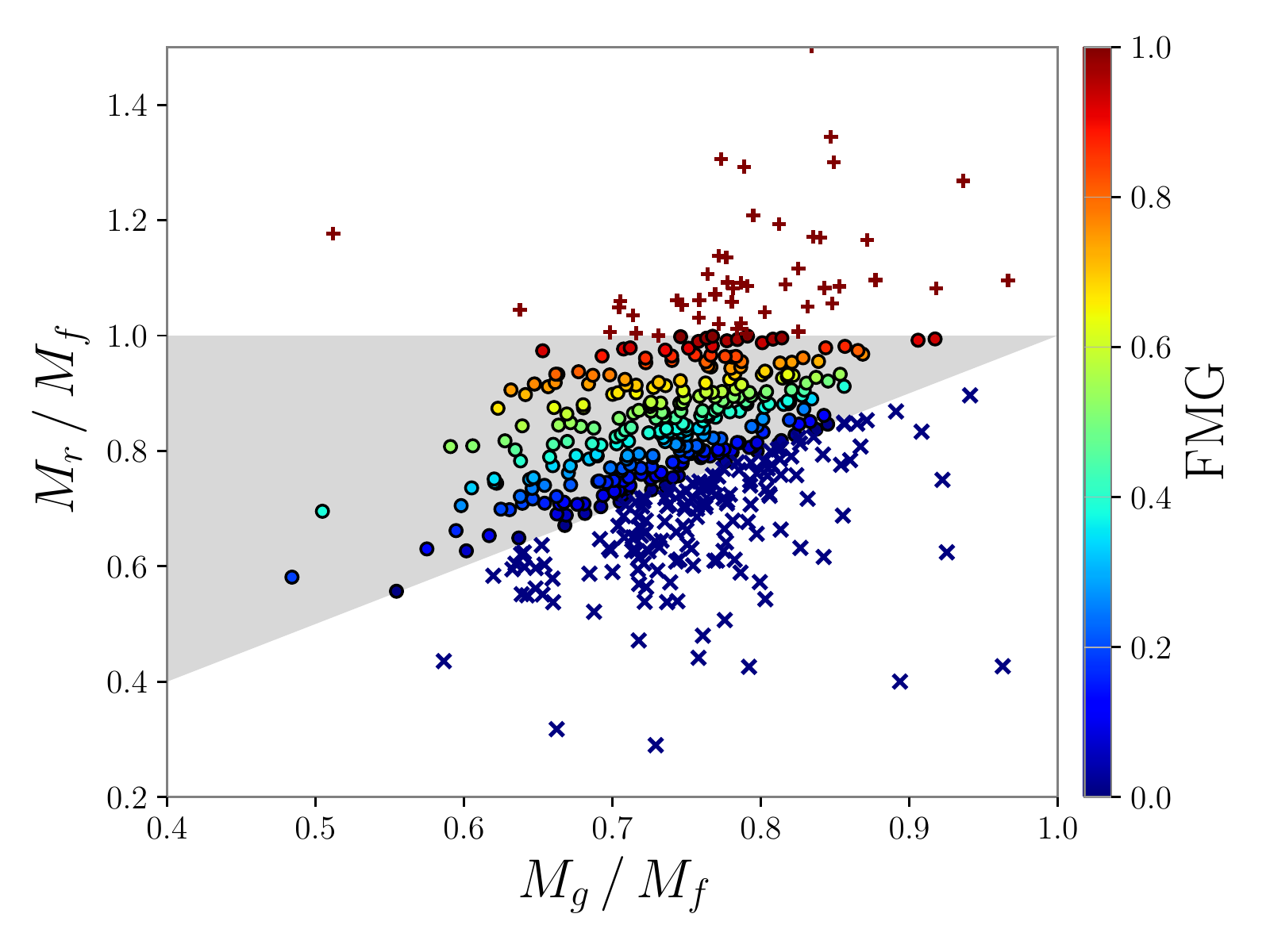}
    \caption{Scatterplot of $M_r/M_f$ vs.\ $M_g/M_f$ for the exploration set with $\rho_N=11.3$, $(m_1,m_2)=(20,15)$ \(M_\odot\), $t_d = 130$ ms and $t_e = 30$ ms. The circles denote samples with $0<\text{\ac{FMG}}<1$, the $\times$ markers denote samples with \ac{FMG} $\le 0$ and the $+$ markers denote overfitted samples with \ac{FMG} $>1$. The 
gray area denotes the region of the parameter space with $0<\text{\ac{FMG}}\le 1$, which contains 59\% of the reconstructed time series.}
    \label{fig:snr8_m1_20_m2_15_tbm_30ms_dur_130ms_roc_O_r_O_g}
\end{figure}

\begin{figure}[ht!]
    \centering
    \includegraphics[width=0.7\linewidth]{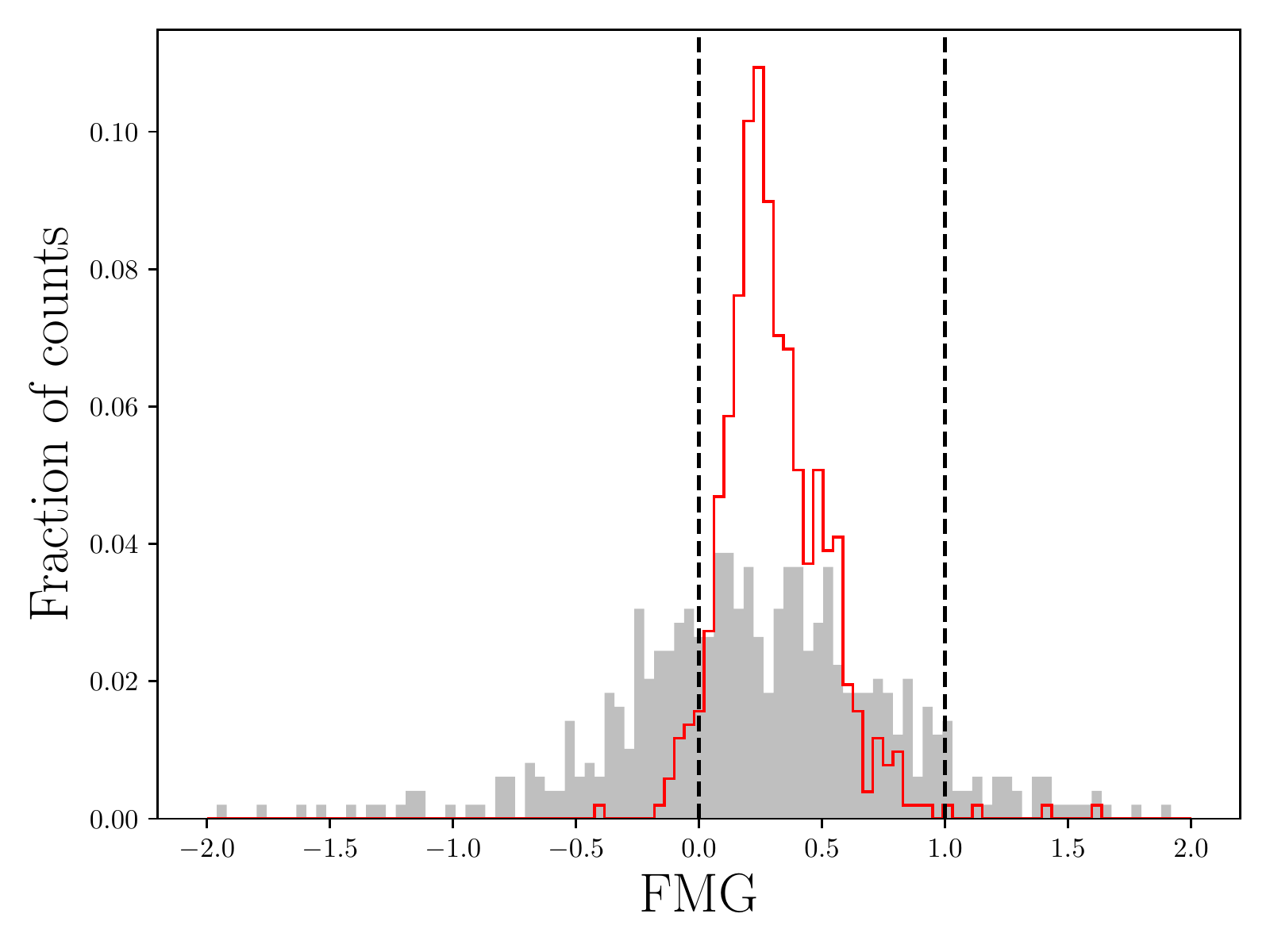}
    \caption{Distribution of the \ac{FMG} for the exploration sets with 
      component masses $(m_1,m_2)=(20,15)$ \(M_\odot\), gate duration $t_d = 130$ ms, gate end-time $t_e = 30$ ms, and $\rho_N=11.3$ (gray-filled) and $\rho_N=42.4$ (red). The vertical dashed lines denote \ac{FMG}$=0$ and \ac{FMG}$=1$. The efficiency of the set with $\rho_N=11.3$ is 59\%. The efficiency of the set with $\rho_N=42.3$ is 95\%.}
    \label{fig:hist_lambda}
\end{figure}

The \ac{FMG} assesses how well the NNETFIX reconstructed data matches the signal in comparison to the full data and gated data.
Positive (negative) values of the \ac{FMG} correspond to $M_r$ greater (smaller) than $M_g$, indicating that the NNETFIX reconstructed time series has a better (worse) match with the
injected waveform than the gated time series. Values of the \ac{FMG} larger than 1 indicate that the \ac{ANN} overfits the data, i.e., the reconstructed time series is more similar to the
injected waveform than the full time series. Therefore, we consider the reconstructions with $0<\text{\ac{FMG}}\le 1$ to be successful. Figures
\ref{fig:snr30_m1_20_m2_15_tbm_30ms_dur_130ms_roc_O_r_O_g} and \ref{fig:snr8_m1_20_m2_15_tbm_30ms_dur_130ms_roc_O_r_O_g} show the distributions of the \ac{FMG} for two exploration sets from
the medium-mass scenario. Figure \ref{fig:hist_lambda} displays the comparison of these distributions.

We quantify NNETFIX's performance by estimating the reconstruction efficiency, which we define as the
fraction of successfully reconstructed samples, i.e., samples with $0<\text{\ac{FMG}}\leq 1$.
The fractions of samples with \ac{FMG} $\le 0$, $0<\text{\ac{FMG}}\leq 1$ and \ac{FMG}
$>1$ for all exploration sets are given in Tables \ref{table:m1_12_m2_10}-\ref{table:m1_35_m2_29}.
The efficiency across all exploration sets varies from approximately $0.31$ to over $0.95$. There is a mild dependence on the component masses of the system; the median value of the efficiency decreases from $0.77$ for the
low-mass scenario to $0.61$ for the high-mass scenario when all other parameters (\ac{SNR}, gate duration and gate end-time) are held fixed.

Within each mass scenario when the gate duration and gate end-time are held fixed, NNETFIX's efficiency typically improves by a factor $\sim 1.5$--2 as the network \ac{SNR} increases.
As the \ac{SNR} becomes higher,
the algorithm can rely  on a larger amount of signal energy before and after the gated portion of the data to reconstruct the time series.  NNETFIX successfully reconstructs over two
thirds of the time series with $\rho_N=28.3$ or larger for all low-mass and medium-mass exploration sets and over half of the time series for the high-mass sets with the exception of two
marginal cases with gate duration $t_d=75$ ms and gate end-time $t_e=90$ ms. The exploration sets with $\rho_N=11.3$ exhibit lower efficiencies, ranging from 31\% for the high-mass set with
$t_d=75$ ms and $t_e=90$ ms to 66\% for the low-mass set with $t_d=130$ ms and $t_e=15$ ms.

\begin{figure}[ht!]
    \centering
    \includegraphics[width=0.7\linewidth]{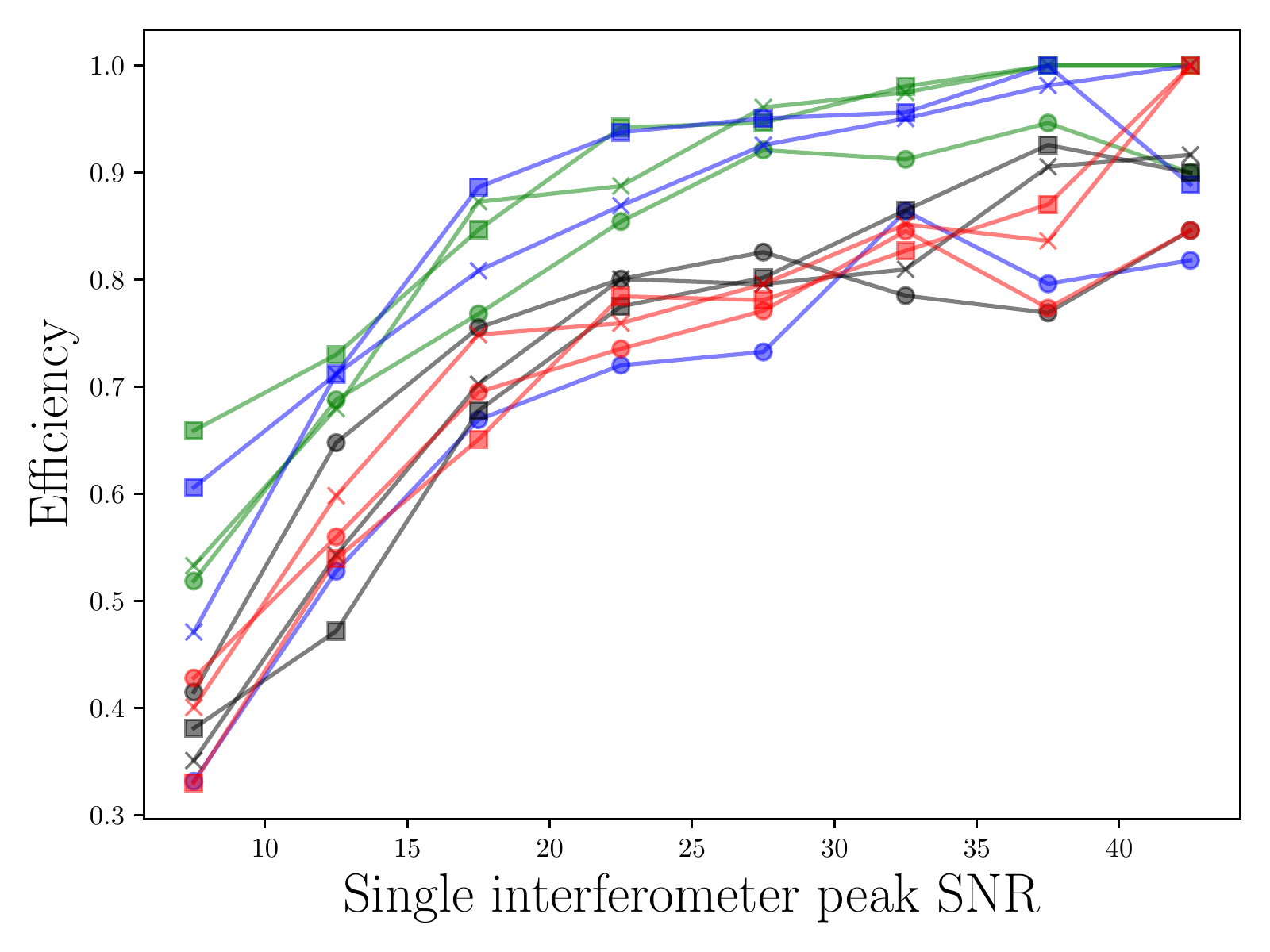}
    \caption{Efficiency as a function of the single interferometer peak \ac{SNR} for the scenario with component masses $(m_1,m_2)=(20,15)$ \(M_\odot\). Each line corresponds to a different gate duration and gate end-time combination. Green (blue, black, red) markers denote gate end-times $t_e = 15$ (30, 90, 170) ms. Circles (crosses, squares) denote gate durations $t_d = 50$ (75, 130) ms. The bin width is 5.}
    \label{fig:frac_sample_m1_20_m2_15}
\end{figure}

Figure \ref{fig:frac_sample_m1_20_m2_15} shows the efficiency for the exploration sets with component masses $(m_1,m_2) = (20,15) M_\odot$ as a function of the single interferometer peak
\ac{SNR}. The percentage of successful reconstructions ranges from $\sim$ 33\%--66\% at low peak \ac{SNR} to $\gtrsim$ 80\% at high peak \ac{SNR}, with the lowest values $\lesssim$ 40\% occurring for
the sets with $t_d\ge 75$ ms and $t_e\ge 30$ ms. Time series with peak \ac{SNR} above $\sim$ 20 show successful reconstructions in 70\% or more of the cases, irrespective of gate duration and
end-time. 

Changing the gate duration does not seem to have a significant effect on NNETFIX's efficiency, which only varies slightly at fixed network \ac{SNR} and gate end-time across all exploration sets.
Similarly, for fixed gate duration and network \ac{SNR}, the gate end-time before merger time also has a marginal effect, although NNETFIX tends to produce better reconstructions when the gate is closer to
the merger time, especially for long gate durations in the low-mass and medium-mass scenarios.

In conclusion, we find that NNETFIX may successfully reconstruct gated data of durations up to a few hundreds of milliseconds and as close as a few tens of milliseconds before the merger
time for a majority of time series with single interferometer peak \ac{SNR} greater than 20.

%% file: effect_skylocalization.tex
\section{Performance of sky maps}\label{effect_skylocalization}

The NNETFIX reconstructed time series are expected to produce better sky maps, and therefore better sky localization error regions of the astrophysical signal, than the gated time series. We evaluate this improvement by comparing
the overlaps of the sky map derived from the full time series with the sky maps
derived from the gated time series and the reconstructed time series. In the following, we generate the sky maps with a modified version of a pyCBC \cite{alex_nitz_2020_3904502} script, \textsc{pycbc\_make\_skymap}, in which the data can be manually
gated.

We follow Ref.~\cite{Ashton_2018} and define the overlap of two sky maps (1,2) as
\begin{equation}\label{eq:def_overlap_integral}
    O_{1,2} = \frac{\displaystyle 4\pi\int \! p_1(\Omega) p_2(\Omega) \, \mathrm{d}\Omega}
    {\displaystyle \int \! p_1(\Omega) \, \mathrm{d}\Omega
    \int \! p_2(\Omega) \, \mathrm{d}\Omega}\,,
\end{equation}
where $p_1(\Omega)$ and $p_2(\Omega)$ are the sky localization probability densities of the sky maps and the integrals are over the solid angle $\Omega$. The discretized version of
Eq.~(\ref{eq:def_overlap_integral}) is
\begin{equation}\label{eq:def_overlap_discrete}
    O_{1,2} = N \sum_{i=1}^N P_{1i}P_{2i}\,,
\end{equation}
where $P_{1i}$ and $P_{2i}$ each denote the sky localization probability of the $i$-th pixel of the corresponding sky map, and $N$ is the total number of pixels.
Each sky map is normalized such that the
sum of the pixel values over the entire map is 1.
Equation (\ref{eq:def_overlap_discrete}) gives values in the range $(0,N)$. Higher values of $O_{1,2}$ indicate a better overlap between the two maps while lower values denote worse overlaps and/or maps which tend to have less-localized error regions.

A suitable metric to evaluate the improvement in the sky localization of a signal due to NNETFIX's reconstruction is the \ac{ORL}
\begin{equation}\label{eq:log_ratio}
    \text{\ac{ORL}} = \log_{10}\frac{O_{r,f}}{O_{g,f}}\,,
\end{equation}
where $O_{r,f}$ ($O_{g,f}$) denote the overlaps of the sky maps obtained with the reconstructed (gated) time series and the full series. Positive (negative) values of the \ac{ORL} indicate
that the sky map from the reconstructed time series has a larger (smaller) overlap with the sky map from the full time series than the latter has with the sky
map from the gated time series. Tables \ref{table:m1_12_m2_10_OI}-\ref{table:m1_35_m2_29_OI} give the fraction of samples with positive \ac{ORL} for all exploration sets. 

High values of the \ac{ORL} are obtained when the overlap of the reconstructed (gated) sky map with the sky map from the full time series is large (small). The former typically
occurs for reconstructed time series with large values of the \ac{FMG}. The latter may happen when the loss of signal due to the gate is high and even small gains in the single interferometer peak \ac{SNR} have a significant impact on the sky localization.

\begin{figure}[ht!]
    \centering
    \includegraphics[width=0.7\linewidth]{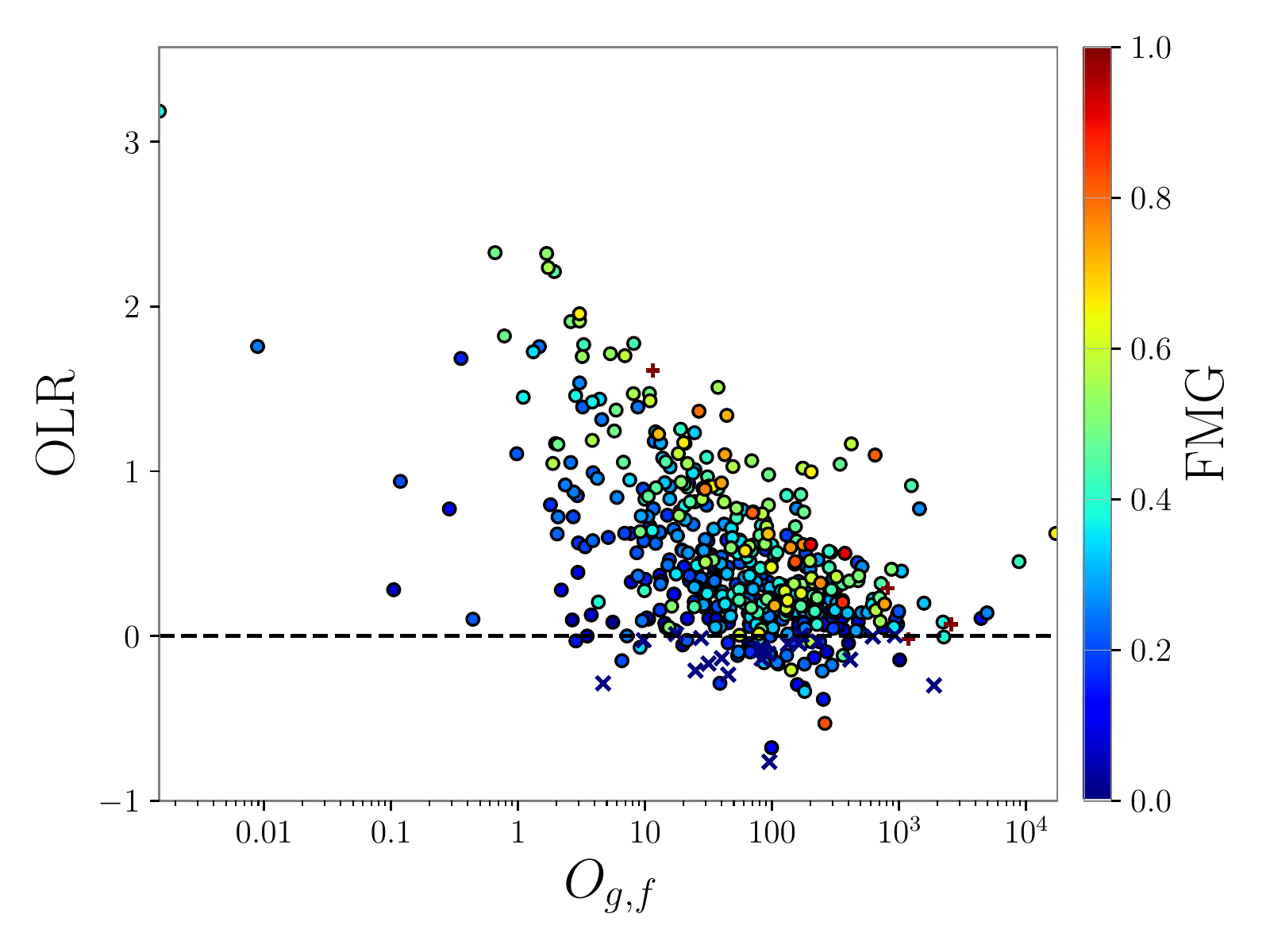}
    \caption{Scatterplot of \ac{ORL} and $O_{g,f}$ for the 512 samples from the exploration set with $\rho_N=42.4$, component masses $(m_1,m_2)=(20,15)$ \(M_\odot\), gate end-time $t_e = 30$ ms and gate duration $t_d = 130$ ms. The colored circles denote samples with $0 < \text{\ac{FMG}} \le 1$, the $\times$ markers denote samples with \ac{FMG} $\le 0$ and the $+$ markers denote overfitted samples with \ac{FMG}  $> 1$. 87\% of the samples have positive \ac{ORL}. The median value is \ac{ORL}$=1.14^{+1.15}_{-1.10}$, where the error is a 1-$\sigma$ percentile.}
    \label{fig:snr30_m1_20_m2_15_tbm_30ms_dur_130ms_roc_OI}
\end{figure}

An example of the \ac{ORL} distribution as a function of $O_{g,f}$ is shown in Fig.\ \ref{fig:snr30_m1_20_m2_15_tbm_30ms_dur_130ms_roc_OI} for the exploration set with $\rho_{N}=42.4$,
component masses $(m_1,m_2)=(20,15)$ $M_\odot$, $t_g=130$ ms and $t_e=30$ ms. The median value of the \ac{ORL} is $1.14^{+1.15}_{-1.10}$, where the error is a 1-$\sigma$ percentile.
The \ac{ORL} is positive in 87\% of the samples.  Median values of \ac{ORL} for all exploration sets are given in Tables \ref{table:m1_12_m2_10_med_Z}-\ref{table:m1_35_m2_29_med_Z}. 

Values of \ac{ORL} across the exploration sets generally increase with network \ac{SNR}, component masses and gate duration. The network \ac{SNR} of the signal is the main factor that determines the value of
the \ac{ORL}. Because NNETFIX efficiently reconstructs time series containing signals with large \acp{SNR}, when network \acp{SNR} are large, the sky maps obtained from the full data are typically more similar to the sky maps from the
reconstructed data than to the sky maps from the gated data.
We find positive \ac{ORL} median values for all exploration sets with $\rho_N\ge 28.3$, irrespective of mass, gate duration and
end-time. For these sets, the median values of the \ac{ORL} for the high \ac{SNR} sets are greater than the corresponding values for the medium \ac{SNR} sets by a factor ranging from $\sim
1.4$ for the high-mass scenario with $t_d=130$ ms and $t_e=15$ ms to $\sim 5$ for the low-mass scenario with $t_d=75$ ms and $t_e=170$ ms. The sky maps of reconstructed time series
with lower \ac{SNR} generally show little improvement compared with the sky maps of gated time series. Median values of \ac{ORL} for the sets with $\rho_{N}=11.3$ are typically around
0, irrespective of the mass scenario, gate duration and gate end-time.

The second most important factor that determines the \ac{ORL} are the component masses. Median \ac{ORL} values typically increase as values of the component masses become larger. For the exploration
sets with $\rho_{N}=28.3$ $(42.4)$, the median values of \ac{ORL} for the high-mass exploration sets are greater than the corresponding values for the low-mass exploration sets by a factor ranging from
$\sim 1.5$ ($2.3$) for $t_d=130$ ms and $t_e=90$ ms ($t_d=130$ ms and $t_e=15$ ms) to $\sim 20.5$ ($12.6$) for $t_d=50$ ms and $t_e=15$ ms ($t_d=50$ ms and $t_e=30$ ms).

Median values of \ac{ORL} have a roughly linear dependency on gate duration. For the high and medium \ac{SNR} exploration sets, the median values of \ac{ORL} for $t_d = 130$ ms are larger than the
corresponding values for $t_d=50$ ms by a factor ranging from $\sim 1.6$  (medium-mass scenario with $\rho_N=42.4$ and $t_e=15$ ms) to $\sim 4.7$ (high-mass scenario with $\rho_N=28.3$
and $t_e=90$ ms).  Since longer gate durations correspond to greater signal losses, NNETFIX's reconstruction provides larger \ac{SNR} gains and \ac{ORL} values as the gate duration
increases. 

\begin{figure}[ht!]
    \centering
    \includegraphics[width=0.7\linewidth]{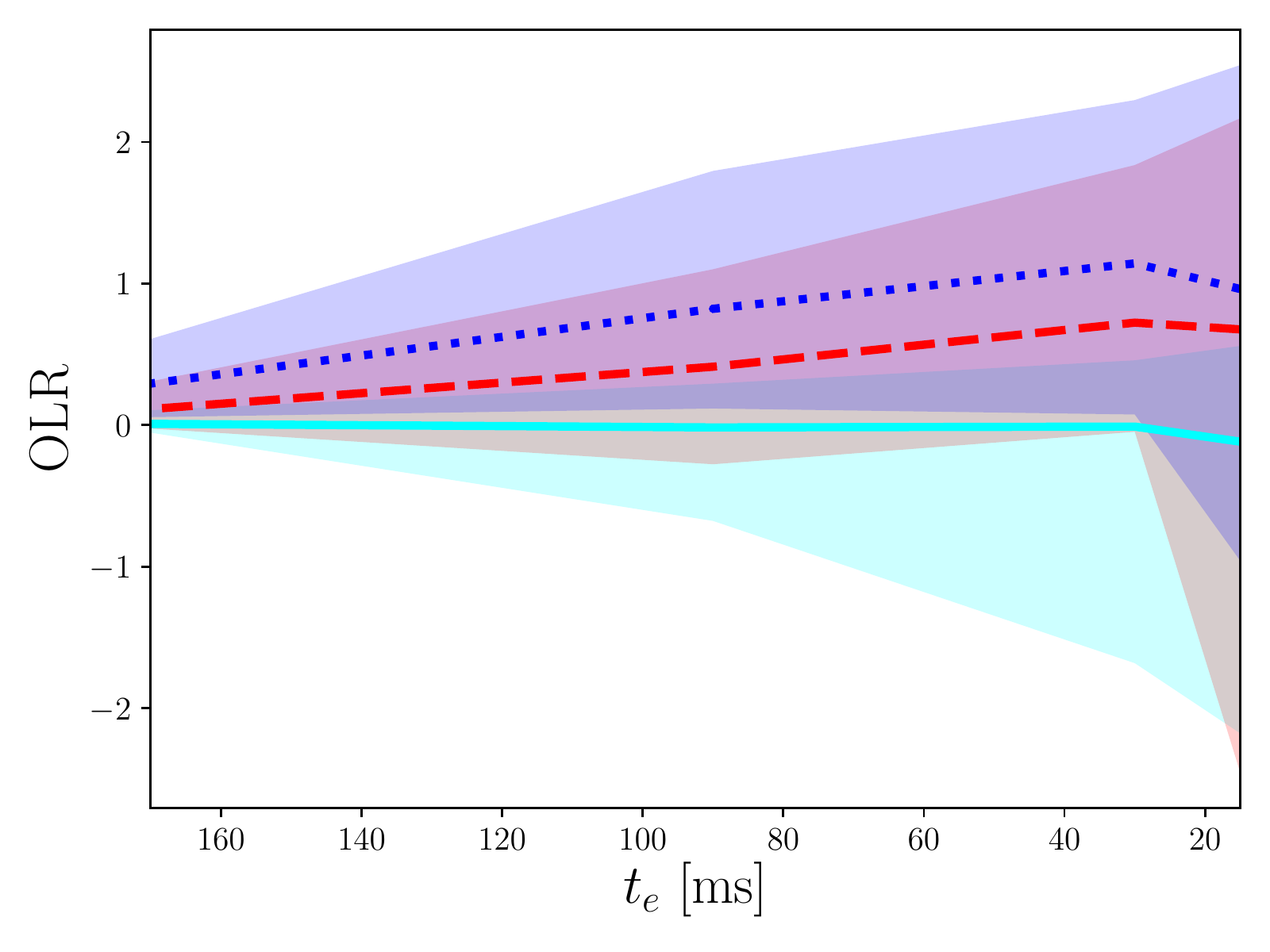}
    \caption{\ac{ORL} as a function of the gate end-time $t_e$ for the exploration set with component masses $(m_1,m_2)=(35,29)$ $M_\odot$, gate duration $t_d = 130$ ms,
      and different network \acp{SNR} $\rho_N=11.3$ (cyan solid), $\rho_N=28.3$ (red dashed), and $\rho_N=42.4$ (blue dashed).
      The curves denote median values. Shaded areas are 1-$\sigma$ percentiles.}
    \label{fig:Z_VS_tbm}
\end{figure}

\begin{figure}[ht!]
    \centering
    \includegraphics[width=0.7\linewidth]{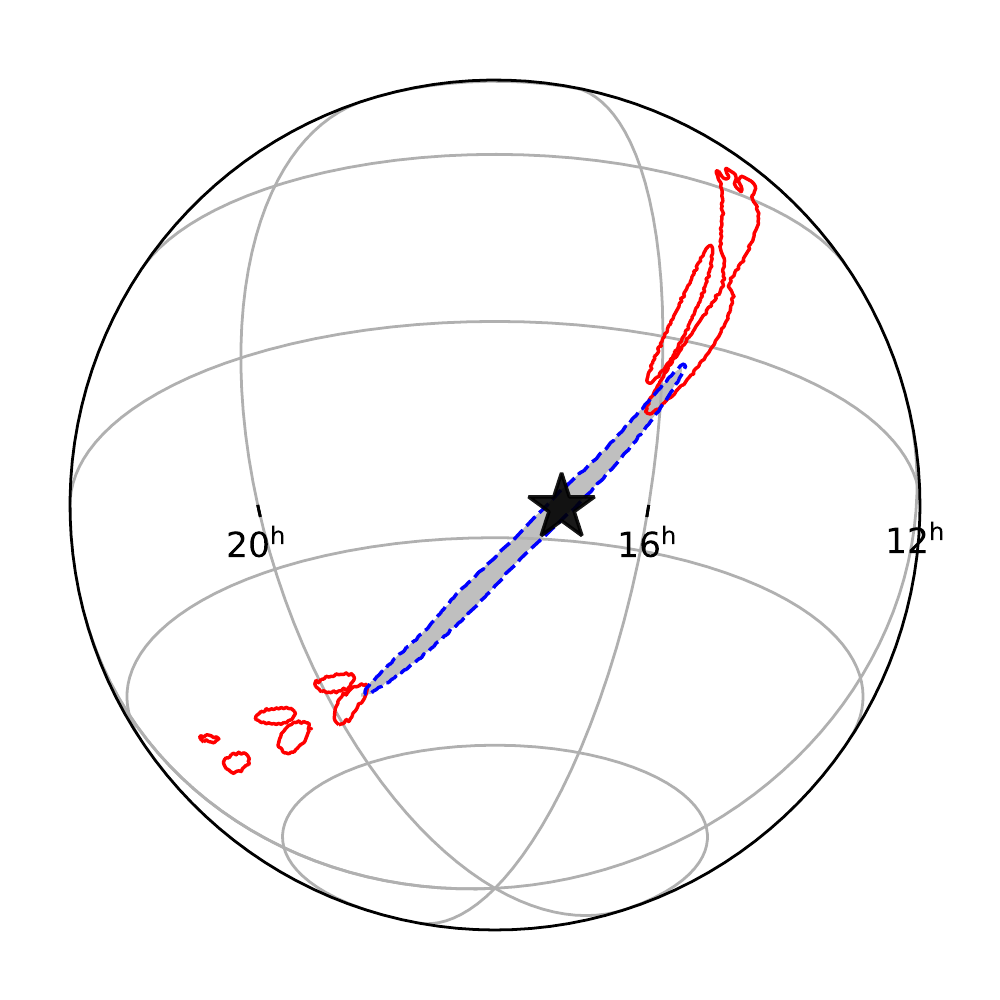}
    \caption{The 90\% probability sky localization error regions obtained with the reconstructed (dashed-blue), full (gray area) and gated (solid-red) time series for the case of Fig.~\ref{fig:inj_result}. The star denotes the injection location.}
    \label{fig:skymap}
\end{figure}

The portion of a signal close to the merger time has a greater impact on the sky map than the portion of the signal in the early inspiral phase.
Therefore, median values of the \ac{ORL} for the medium and high \ac{SNR} exploration sets with $t_e = 15$ ms are typically higher than the corresponding values for the sets with $t_e=170$ ms by a
factor ranging from $\sim 1.1$ (low-mass scenario with $\rho_N=28.3$ and $t_d=90$ ms) to $\sim 7.1$ (high-mass scenario with $\rho_N=28.3$ and $t_d=50$ ms). For shorter signals and larger
gate durations, a gate end-time very close to the merger time may lead to large signal losses and removal of the merger portion of the signal in \ac{H1},  and thus, make the reconstruction
process less efficient. Figure \ref{fig:Z_VS_tbm} shows the \ac{ORL} as a function of the gate end-time for the high-mass sets with $t_d=130$ ms and different network \acp{SNR}. Figure
\ref{fig:skymap} shows the sky localization error region that is obtained with the NNETFIX reconstructed data for the case of Fig.~\ref{fig:inj_result}. The value of the \ac{ORL} is $\sim 1.7$,
corresponding to improving the overlap by a factor of $\sim 50$.

In summary, for a majority of the cases with gate durations up to a few hundreds of milliseconds and as close as a few tens of milliseconds to the merger time, the sky maps of
reconstructed time series with network \ac{SNR} $\rho_N \ge 28.3$ better overlap with the sky maps of the full time series compared to the sky maps
obtained with gated data (in some cases by a factor up to over 1000\%).
In these cases, it can also be shown that the true direction of the signals typically belong to sky localization error regions for the reconstructed data with
smaller probability contour values than the regions obtained with gated data.

%% file: conclusion.tex
\section{Conclusion} \label{conclusion}

In this paper, we have presented NNETFIX, a new machine learning-based algorithm designed to estimate the portion of a \ac{BBH} \ac{GW} signal that may be gated due to the
presence of an overlapping glitch.
We have tested the accuracy of the algorithm with different choices of signal parameters and gate settings,
and defined several metrics to assess NNETFIX's performance.
Among these metrics, the most important ones are the \ac{FMG} and the \ac{ORL}.

The \ac{FMG} quantifies the
algorithm's efficiency in reconstructing the gated data in the time domain.
Positive values of this metric indicate that the full time series better matches the NNETFIX reconstructed time series than the gated time series.
The fraction of samples that show improvement varies from approximately one
third to over 95\% across the cases that we investigated.
Results show that NNETFIX may be able to successfully reconstruct a majority of \ac{BBH} signals with peak single interferometer
\ac{SNR} greater than 20 and gates with durations up to a few hundreds of milliseconds as close as a few tens of milliseconds before their merger time.  

The \ac{ORL} quantifies the algorithm's efficiency in improving the sky map from the gated time series. Positive values of this metric indicate that the sky map from the
NNETFIX reconstructed time series has a larger overlap with the sky map from the full time series than the one obtained from the gated data. Sky maps
from reconstructed data improve for higher network \ac{SNR} values as the \ac{ANN} can use a larger amount of signal energy to estimate the missing portion of the waveform. We find positive \ac{ORL}
median values for all cases that we investigated with network \ac{SNR} above 28.3. Perhaps surprisingly, NNETFIX seems also to perform better in cases with longer gate
durations or shorter signals. In these scenarios, the sky localizations obtained with gated data are considerably degraded. Thus, the improvements in the reconstructed sky maps
are more sizeable. Reconstructed sky maps of more massive \ac{BBH} mergers typically show significant improvements compared to the sky maps obtained with gated data.

In a real case scenario, we envision NNETFIX to be pre-trained on real noise data from the detectors for a sparse set of models, each covering a region of the five-dimensional
parameter space spanned by \ac{SNR}, component masses, gate duration, and gate end-time before geocentric merger time, as was illustrated in Sec.~\ref{methodology} (but with a
finer grid). While the optimal value of the network \ac{SNR} and the best estimates of the signal component masses are unknown to the observer because of the gating, the gated
data and/or the data from the second interferometer may provide a rough estimate of these parameters.
The estimated values of these parameters
and the known gate parameters
can then be used to choose the most appropriate pre-trained model in the NNETFIX bank. Typically, the NNETFIX reconstructed time series will produce a higher
single interferometer peak \ac{SNR} than the gated time series. If that is the case, the known \ac{FSG} can be used to estimate the optimal single interferometer peak \ac{SNR} of
the (unknown) full time series by fitting the expected roughly linear relation between the \ac{FRS} and the \ac{FSG} for the given \ac{TT} set.

The overlap of the sky map from the NNETFIX reconstructed data with the sky map that could be obtained with the full data (if it were not contaminated by a glitch), $O_{r,f}$,
can be estimated by looking at the distribution of the \ac{ORL} for the \ac{TT} set at hand. The exploration sets that we investigated show that there is a well-defined
correlation between the \ac{ORL}, the \ac{FSG} and the overlap between the sky maps obtained from the gated data and the NNETFIX reconstructed data, $O_{r,g}$. If this
correlation is generally valid, the \ac{ORL} can be estimated from the observed values of the $O_{r,g}$ and the \ac{FSG} using a fit calculated from the \ac{TT} set used to train the selected NNETFIX model. To expedite this process, the samples in each \ac{TT} set could be clustered according to the distributions of the \ac{ORL}, the $O_{r,g}$, and the \ac{FSG}. A
classifier could then be trained to estimate the optimal \ac{SNR} and sky localization error region of the full signal.

Once NNETFIX has been trained, the CPU time required to reconstruct the data is of the order of a few seconds for gate durations up to hundreds of milliseconds. This short
turnout makes the algorithm suitable to be used in low-latency. NNETFIX could also be applied to \ac{GW} signals other than \ac{BBH} mergers, such as \ac{BNS} or \ac{NSBH} mergers.
Therefore, it could be beneficial for rapid follow-up of glitch-contaminated, potentially \ac{EM}-bright candidate detections. In future work, we intend to explore the
application of NNETFIX and its effect on the sky localization of \ac{BNS} signals, as well as detector network configurations with more than two detectors. Improving the sky
localizations of potentially \ac{EM}-bright signals could increase the chances of coincident \ac{EM} and \ac{GW} observations and lead to a better understanding of the physical
properties  of their sources.  

%% file: acknowledgements.tex
\section*{Acknowledgements}
K.M., R.Q.J., and M.C.\ are supported by the U.S.\ National Science Foundation grants PHY-1921006 and PHY-2011334. The authors would like to thank their LIGO Scientific Collaboration and
Virgo Collaboration colleagues for their help and useful comments, in particular Tito Dal Canton. The authors are grateful for computational resources provided by the LIGO
Laboratory and supported by the U.S.\ National Science Foundation Grants PHY-0757058 and PHY-0823459, as well as resources from the Gravitational Wave Open Science Center, a service of the
LIGO Laboratory, the LIGO Scientific Collaboration and the Virgo Collaboration. Part of this research has made use of data, software and web tools obtained from the Gravitational Wave Open
Science Center and openly available at \url{https://www.gw-openscience.org}. LIGO was constructed and is operated by the California Institute of Technology and Massachusetts Institute of
Technology with funding from the U.S.\ National Science Foundation under grant PHY-0757058. Virgo is funded by the French Centre National de la Recherche Scientifique (CNRS), the Italian
Istituto Nazionale di Fisica Nucleare (INFN) and the Dutch Nikhef, with contributions by Polish and Hungarian institutes. This manuscript has been assigned LIGO Document Control Center
number LIGO-P2000497.

%% file: evaluation_reconstruction_in_timeseries_tables.tex
\newpage

\newgeometry{top=1.3in, bottom=1.3in}
\begin{landscape}

\begin{table}[h!]
\renewcommand*{\arraystretch}{1.3}
\setlength{\tabcolsep}{4pt}
\begin{minipage}{1.0\linewidth}
\centering
\scalebox{0.7}{
\begin{tabular}{c c|| c c c c c c c c c}
\multicolumn{2}{c }{Network SNR} & \multicolumn{3}{ c }{11.3} & \multicolumn{3}{ c }{28.3} & \multicolumn{3}{ c}{42.4} \\
\cmidrule(rl){1-2}  \cmidrule(rl){3-5} \cmidrule(rl){6-8} \cmidrule(rl){9-11}
\multicolumn{2}{c }{$t_d$ [ms]} &50 & 75 & 130 &50 & 75 & 130 &50 & 75 & 130\\
\midrule\midrule\multirow{4}{*}{$t_e$ [ms]}& 15& 0.33/{\bf 0.54}/0.14& 0.34/{\bf 0.55}/0.11& 0.31/{\bf 0.66}/0.04& 0.21/{\bf 0.76}/0.03& 0.13/{\bf 0.85}/0.01& 0.16/{\bf 0.84}/0.00& 0.10/{\bf 0.88}/0.02& 0.07/{\bf 0.93}/0.01& 0.06/{\bf 0.94}/0.00\\
\cmidrule(rl){2-2} \cmidrule(rl){3-5} \cmidrule(rl){6-8} \cmidrule(rl){9-11}
& 30& 0.30/0.46/0.24& 0.31/0.46/0.23& 0.29/{\bf 0.63}/0.08& 0.15/{\bf 0.78}/0.07& 0.16/{\bf 0.80}/0.04& 0.16/{\bf 0.84}/0.01& 0.06/{\bf 0.88}/0.06& 0.04/{\bf 0.93}/0.04& 0.06/{\bf 0.94}/0.01\\
\cmidrule(rl){2-2} \cmidrule(rl){3-5} \cmidrule(rl){6-8} \cmidrule(rl){9-11}
& 90& 0.28/0.39/0.33& 0.32/0.36/0.32& 0.31/{\bf 0.52}/0.17& 0.10/{\bf 0.69}/0.21& 0.15/{\bf 0.71}/0.13& 0.14/{\bf 0.82}/0.05& 0.03/{\bf 0.80}/0.17& 0.07/{\bf 0.84}/0.09& 0.07/{\bf 0.92}/0.01\\
\cmidrule(rl){2-2} \cmidrule(rl){3-5} \cmidrule(rl){6-8} \cmidrule(rl){9-11}
& 170& 0.28/0.40/0.33& 0.31/0.40/0.29& 0.34/0.38/0.27& 0.09/{\bf 0.68}/0.23& 0.15/{\bf 0.71}/0.14& 0.21/{\bf 0.67}/0.12& 0.02/{\bf 0.84}/0.14& 0.05/{\bf 0.83}/0.13& 0.11/{\bf 0.83}/0.06\\
\bottomrule
\end{tabular}
}
\caption{Fraction of samples with \ac{FMG} $\le 0$, $0 < \text{\ac{FMG}} \le 1$ and \ac{FMG} $>1$ for the exploration sets with component masses $(m_1,m_2)=(12,10) M_\odot$. Boldface entries denote sets where the fraction of samples with $0 < \text{\ac{FMG}} \le 1$ is larger than 50\%.}
\label{table:m1_12_m2_10}
\end{minipage}

\begin{minipage}{1.0\linewidth}
\centering
\scalebox{0.7}{
\begin{tabular}{c c|| c c c c c c c c c}
\multicolumn{2}{c }{Network SNR} & \multicolumn{3}{ c }{11.3} & \multicolumn{3}{ c }{28.3} & \multicolumn{3}{ c }{42.4} \\
\cmidrule(rl){1-2}  \cmidrule(rl){3-5} \cmidrule(rl){6-8} \cmidrule(rl){9-11}
\multicolumn{2}{c }{$t_d$ [ms]} &50 & 75 & 130 &50 & 75 & 130 &50 & 75 & 130\\
\midrule\midrule\multirow{4}{*}{$t_e$ [ms]}& 15& 0.28/{\bf 0.52}/0.19& 0.31/{\bf 0.55}/0.15& 0.29/{\bf 0.64}/0.06& 0.12/{\bf 0.78}/0.10& 0.14/{\bf 0.83}/0.03& 0.13/{\bf 0.87}/0.01& 0.04/{\bf 0.89}/0.06& 0.04/{\bf 0.94}/0.02& 0.05/{\bf 0.95}/0.00\\
\cmidrule(rl){2-2} \cmidrule(rl){3-5} \cmidrule(rl){6-8} \cmidrule(rl){9-11}
& 30& 0.29/0.34/0.37& 0.29/0.49/0.22& 0.31/{\bf 0.59}/0.10& 0.11/{\bf 0.68}/0.21& 0.10/{\bf 0.82}/0.08& 0.11/{\bf 0.88}/0.01& 0.04/{\bf 0.75}/0.21& 0.05/{\bf 0.91}/0.04& 0.04/{\bf 0.95}/0.01\\
\cmidrule(rl){2-2} \cmidrule(rl){3-5} \cmidrule(rl){6-8} \cmidrule(rl){9-11}
& 90& 0.24/0.42/0.34& 0.30/0.38/0.32& 0.29/0.39/0.32& 0.04/{\bf 0.76}/0.20& 0.08/{\bf 0.72}/0.20& 0.17/{\bf 0.69}/0.15& 0.02/{\bf 0.79}/0.20& 0.03/{\bf 0.79}/0.19& 0.10/{\bf 0.80}/0.10\\
\cmidrule(rl){2-2} \cmidrule(rl){3-5} \cmidrule(rl){6-8} \cmidrule(rl){9-11}
& 170& 0.20/0.42/0.37& 0.26/0.42/0.32& 0.31/0.32/0.36& 0.04/{\bf 0.69}/0.27& 0.04/{\bf 0.72}/0.23& 0.11/{\bf 0.69}/0.19& 0.01/{\bf 0.78}/0.21& 0.01/{\bf 0.80}/0.19& 0.06/{\bf 0.79}/0.15\\
\bottomrule
\end{tabular}
}
\caption{Fraction of samples with \ac{FMG} $\le 0$, $0 < \text{\ac{FMG}} \le 1$ and \ac{FMG} $>1$ for the exploration sets with component masses $(m_1,m_2)=(20,15) M_\odot$. Boldface entries denote sets where the fraction of samples with $0 < \text{\ac{FMG}} \le 1$ is larger than 50\%.}\label{table:m1_20_m2_15}
\end{minipage}
\begin{minipage}{1.0\linewidth}
\centering
\scalebox{0.7}{
\begin{tabular}{c c|| c c c c c c c c c}
\multicolumn{2}{c }{Network SNR} & \multicolumn{3}{ c }{11.3} & \multicolumn{3}{ c }{28.3} & \multicolumn{3}{ c }{42.4} \\
\cmidrule(rl){1-2}  \cmidrule(rl){3-5} \cmidrule(rl){6-8} \cmidrule(rl){9-11}
\multicolumn{2}{c }{$t_d$ [ms]} &50 & 75 & 130 &50 & 75 & 130 &50 & 75 & 130\\
\midrule\midrule\multirow{4}{*}{$t_e$ [ms]}& 15& 0.19/{\bf 0.50}/0.31& 0.27/{\bf 0.53}/0.20& 0.25/0.44/0.31& 0.04/{\bf 0.76}/0.20& 0.05/{\bf 0.84}/0.11& 0.07/{\bf 0.76}/0.17& 0.02/{\bf 0.80}/0.19& 0.02/{\bf 0.90}/0.08& 0.02/{\bf 0.81}/0.17\\
\cmidrule(rl){2-2} \cmidrule(rl){3-5} \cmidrule(rl){6-8} \cmidrule(rl){9-11}
& 30& 0.21/0.35/0.45& 0.24/0.34/0.42& 0.24/0.44/0.32& 0.03/{\bf 0.60}/0.37& 0.03/{\bf 0.56}/0.41& 0.05/{\bf 0.76}/0.19& 0.01/{\bf 0.62}/0.37& 0.03/{\bf 0.66}/0.31& 0.02/{\bf 0.86}/0.12\\
\cmidrule(rl){2-2} \cmidrule(rl){3-5} \cmidrule(rl){6-8} \cmidrule(rl){9-11}
& 90& 0.17/0.47/0.36& 0.12/0.31/0.57& 0.21/0.32/0.47& 0.01/{\bf 0.75}/0.24& 0.02/0.42/0.57& 0.04/{\bf 0.51}/0.46& 0.00/{\bf 0.79}/0.21& 0.00/0.38/0.62& 0.01/{\bf 0.59}/0.40\\
\cmidrule(rl){2-2} \cmidrule(rl){3-5} \cmidrule(rl){6-8} \cmidrule(rl){9-11}
& 170& 0.14/0.42/0.44& 0.15/0.40/0.45& 0.22/0.45/0.33& 0.01/{\bf 0.63}/0.36& 0.01/{\bf 0.64}/0.35& 0.05/{\bf 0.75}/0.20& 0.01/{\bf 0.67}/0.32& 0.00/{\bf 0.65}/0.35& 0.01/{\bf 0.88}/0.11\\
\bottomrule
\end{tabular}
}
\caption{Fraction of samples with \ac{FMG} $\le 0$, $0 < \text{\ac{FMG}} \le 1$ and \ac{FMG} $>1$ for the exploration sets with component masses $(m_1,m_2)=(35,29) M_\odot$. Boldface entries denote sets where the fraction of samples with $0 < \text{\ac{FMG}} \le 1$ is larger than 50\%.}\label{table:m1_35_m2_29}
\end{minipage}

\end{table}

\end{landscape}
\restoregeometry

%% file: effect_skylocalization_tables.tex
\begin{table}[h!]
\renewcommand*{\arraystretch}{1.3}
\setlength{\tabcolsep}{4pt}
\begin{minipage}{1.0\linewidth}
\centering
\scalebox{0.85}{
\begin{tabular}{c c|| c c c c c c c c c}
\multicolumn{2}{ c }{Network SNR} & \multicolumn{3}{ c }{11.3} & \multicolumn{3}{ c }{28.3} & \multicolumn{3}{ c }{42.4} \\
\cmidrule(rl){1-2}  \cmidrule(rl){3-5} \cmidrule(rl){6-8} \cmidrule(rl){9-11}
\multicolumn{2}{ c }{$t_d$ [ms]} &50 & 75 & 130 &50 & 75 & 130 &50 & 75 & 130\\
\midrule\midrule\multirow{4}{*}{$t_e$ [ms]}& 15& ${0.52}$& ${0.53}$& ${0.56}$& ${0.62}$& ${0.71}$& ${0.74}$& ${0.76}$& ${0.83}$& ${0.87}$\\
\cmidrule(rl){2-2} \cmidrule(rl){3-5} \cmidrule(rl){6-8} \cmidrule(rl){9-11}
& 30& ${0.56}$& ${0.53}$& ${0.52}$& ${0.68}$& ${0.73}$& ${0.76}$& ${0.81}$& ${0.85}$& ${0.91}$\\
\cmidrule(rl){2-2} \cmidrule(rl){3-5} \cmidrule(rl){6-8} \cmidrule(rl){9-11}
& 90& ${0.55}$& ${0.51}$& ${0.53}$& ${0.69}$& ${0.65}$& ${0.74}$& ${0.78}$& ${0.79}$& ${0.86}$\\
\cmidrule(rl){2-2} \cmidrule(rl){3-5} \cmidrule(rl){6-8} \cmidrule(rl){9-11}
& 170& ${0.61}$& ${0.58}$& ${0.51}$& ${0.72}$& ${0.64}$& ${0.65}$& ${0.79}$& ${0.77}$& ${0.76}$\\
\bottomrule
\end{tabular}
}
\caption{Fraction of samples with positive \ac{ORL} for the exploration sets with component masses $(m_1,m_2)=(12,10) M_\odot$.}
\label{table:m1_12_m2_10_OI}
\end{minipage}

\vskip 24pt

\begin{minipage}{1.0\linewidth}
\centering
\scalebox{0.85}{
\begin{tabular}{c c|| c c c c c c c c c} \toprule
\multicolumn{2}{ c }{Network SNR} & \multicolumn{3}{ c }{11.3} & \multicolumn{3}{ c }{28.3} & \multicolumn{3}{ c }{42.4} \\
\cmidrule(rl){1-2}  \cmidrule(rl){3-5} \cmidrule(rl){6-8} \cmidrule(rl){9-11}
\multicolumn{2}{ c }{$t_d$ [ms]} &50 & 75 & 130 &50 & 75 & 130 &50 & 75 & 130\\
\midrule\midrule\multirow{4}{*}{$t_e$ [ms]}& 15& ${0.51}$& $\textit{0.49}$& ${0.55}$& ${0.68}$& ${0.74}$& ${0.79}$& ${0.84}$& ${0.85}$& ${0.84}$\\
\cmidrule(rl){2-2} \cmidrule(rl){3-5} \cmidrule(rl){6-8} \cmidrule(rl){9-11}
& 30& $\textit{0.46}$& $\textit{0.48}$& ${0.51}$& ${0.69}$& ${0.76}$& ${0.80}$& ${0.85}$& ${0.88}$& ${0.88}$\\
\cmidrule(rl){2-2} \cmidrule(rl){3-5} \cmidrule(rl){6-8} \cmidrule(rl){9-11}
& 90& ${0.61}$& ${0.50}$& $\textit{0.47}$& ${0.74}$& ${0.75}$& ${0.65}$& ${0.86}$& ${0.85}$& ${0.79}$\\
\cmidrule(rl){2-2} \cmidrule(rl){3-5} \cmidrule(rl){6-8} \cmidrule(rl){9-11}
& 170& ${0.62}$& ${0.56}$& $\textit{0.47}$& ${0.77}$& ${0.78}$& ${0.67}$& ${0.84}$& ${0.87}$& ${0.82}$\\
\bottomrule
\end{tabular}
}
\caption{Fraction of samples with positive \ac{ORL} for the exploration sets with component masses $(m_1,m_2)=(20,15) M_\odot$. Entries in italic denote sets where the fraction of samples is smaller than 0.5.}
\label{table:m1_20_m2_15_OI}
\end{minipage}

\vskip 24pt

\begin{minipage}{1.0\linewidth}
\centering
\scalebox{0.85}{
\begin{tabular}{c c|| c c c c c c c c c} \toprule
\multicolumn{2}{ c }{Network SNR} & \multicolumn{3}{ c }{11.3} & \multicolumn{3}{ c }{28.3} & \multicolumn{3}{ c }{42.4} \\
\cmidrule(rl){1-2}  \cmidrule(rl){3-5} \cmidrule(rl){6-8} \cmidrule(rl){9-11}
\multicolumn{2}{ c }{$t_d$ [ms]} &50 & 75 & 130 &50 & 75 & 130 &50 & 75 & 130\\
\midrule\midrule\multirow{4}{*}{$t_e$ [ms]}& 15& ${0.57}$& ${0.53}$& $\textit{0.43}$& ${0.77}$& ${0.79}$& ${0.69}$& ${0.85}$& ${0.83}$& ${0.73}$\\
\cmidrule(rl){2-2} \cmidrule(rl){3-5} \cmidrule(rl){6-8} \cmidrule(rl){9-11}
& 30& ${0.51}$& $\textit{0.45}$& $\textit{0.49}$& ${0.81}$& ${0.81}$& ${0.82}$& ${0.91}$& ${0.87}$& ${0.87}$\\
\cmidrule(rl){2-2} \cmidrule(rl){3-5} \cmidrule(rl){6-8} \cmidrule(rl){9-11}
& 90& ${0.65}$& ${0.63}$& $\textit{0.46}$& ${0.84}$& ${0.80}$& ${0.75}$& ${0.93}$& ${0.89}$& ${0.87}$\\
\cmidrule(rl){2-2} \cmidrule(rl){3-5} \cmidrule(rl){6-8} \cmidrule(rl){9-11}
& 170& ${0.65}$& ${0.64}$& ${0.57}$& ${0.79}$& ${0.84}$& ${0.80}$& ${0.88}$& ${0.91}$& ${0.91}$\\
\bottomrule
\end{tabular}
}
\caption{Fraction of samples with positive \ac{ORL} for the exploration sets with component masses $(m_1,m_2)=(35,29) M_\odot$. Entries in italic denote sets where the fraction of samples is smaller than 0.5.}
\label{table:m1_35_m2_29_OI}
\end{minipage}
\end{table}

%%%%%%  for median of Z %%%%%%%%%%%%%%%%%
\begin{table}[h!]

\renewcommand*{\arraystretch}{1.3}
\setlength{\tabcolsep}{4pt}
\begin{minipage}{1.0\linewidth}
\centering
\scalebox{0.85}{
\begin{tabular}{c c|| l l l l l l l l l}
\multicolumn{2}{ c }{Network SNR} & \multicolumn{3}{ c }{11.3} & \multicolumn{3}{ c }{28.3} & \multicolumn{3}{ c }{42.4} \\
\cmidrule(rl){1-2}  \cmidrule(rl){3-5} \cmidrule(rl){6-8} \cmidrule(rl){9-11}
\multicolumn{2}{ c }{$t_d$ [ms]} &50 & 75 & 130 &50 & 75 & 130 &50 & 75 & 130\\
\midrule\midrule\multirow{4}{*}{$t_e$ [ms]}& 15& ${0.0012}$& ${0.0064}$& ${0.0096}$& ${0.016}$& ${0.031}$& ${0.056}$& ${0.051}$& ${0.086}$& ${0.14}$\\
\cmidrule(rl){2-2} \cmidrule(rl){3-5} \cmidrule(rl){6-8} \cmidrule(rl){9-11}
& 30& ${0.0044}$& ${0.0059}$& ${0.0022}$& ${0.015}$& ${0.028}$& ${0.044}$& ${0.044}$& ${0.073}$& ${0.14}$\\
\cmidrule(rl){2-2} \cmidrule(rl){3-5} \cmidrule(rl){6-8} \cmidrule(rl){9-11}
& 90& ${0.0032}$& ${0.0005}$& ${0.0021}$& ${0.0091}$& ${0.011}$& ${0.034}$& ${0.026}$& ${0.04}$& ${0.10}$\\
\cmidrule(rl){2-2} \cmidrule(rl){3-5} \cmidrule(rl){6-8} \cmidrule(rl){9-11}
& 170& ${0.0026}$& ${0.0037}$& ${0.0007}$& ${0.0076}$& ${0.0063}$& ${0.018}$& ${0.018}$& ${0.032}$& ${0.060}$\\
\bottomrule
\end{tabular}
}
\caption{Median values of \ac{ORL} for the exploration sets with component masses $(m_1,m_2)=(12,10) M_\odot$.}
\label{table:m1_12_m2_10_med_Z}
\end{minipage}

\vskip 24pt

\begin{minipage}{1.0\linewidth}
\centering
\scalebox{0.85}{
\begin{tabular}{c c|| l l l l l l l l l} 
\multicolumn{2}{ c }{Network SNR} & \multicolumn{3}{ c }{11.3} & \multicolumn{3}{ c }{28.3} & \multicolumn{3}{ c }{42.4} \\
\cmidrule(rl){1-2}  \cmidrule(rl){3-5} \cmidrule(rl){6-8} \cmidrule(rl){9-11}
\multicolumn{2}{ c }{$t_d$ [ms]} &50 & 75 & 130 &50 & 75 & 130 &50 & 75 & 130\\
\midrule\midrule\multirow{4}{*}{$t_e$ [ms]}& 15& ${0.0023}$& $\textit{-0.0007}$& ${0.016}$& ${0.054}$& ${0.087}$& ${0.17}$& ${0.16}$& ${0.24}$& ${0.32}$\\
\cmidrule(rl){2-2} \cmidrule(rl){3-5} \cmidrule(rl){6-8} \cmidrule(rl){9-11}
& 30& $\textit{-0.0050}$& $\textit{-0.0029}$& ${0.0039}$& ${0.044}$& ${0.079}$& ${0.15}$& ${0.13}$& ${0.22}$& ${0.31}$\\
\cmidrule(rl){2-2} \cmidrule(rl){3-5} \cmidrule(rl){6-8} \cmidrule(rl){9-11}
& 90& ${0.0049}$& ${0.0002}$& $\textit{-0.0051}$& ${0.025}$& ${0.046}$& ${0.050}$& ${0.074}$& ${0.13}$& ${0.25}$\\
\cmidrule(rl){2-2} \cmidrule(rl){3-5} \cmidrule(rl){6-8} \cmidrule(rl){9-11}
& 170& ${0.0047}$& ${0.0051}$& $\textit{-0.0044}$& ${0.023}$& ${0.037}$& ${0.045}$& ${0.046}$& ${0.098}$& ${0.17}$\\
\bottomrule
\end{tabular}
}
\caption{Median values of \ac{ORL} for the exploration sets with component masses $(m_1,m_2)=(20,15) M_\odot$. Italic entries denote sets with negative values.}
\label{table:m1_20_m2_15_med_Z}
\end{minipage}

\vskip 24pt

\begin{minipage}{1.0\linewidth}
\centering
\scalebox{0.85}{
\begin{tabular}{c c|| l l l l l l l l l} 
\multicolumn{2}{ c }{Network SNR} & \multicolumn{3}{ c }{11.3} & \multicolumn{3}{ c }{28.3} & \multicolumn{3}{ c }{42.4} \\
\cmidrule(rl){1-2}  \cmidrule(rl){3-5} \cmidrule(rl){6-8} \cmidrule(rl){9-11}
\multicolumn{2}{ c }{$t_d$ [ms]} &50 & 75 & 130 &50 & 75 & 130 &50 & 75 & 130\\
\midrule\midrule\multirow{4}{*}{$t_e$ [ms]}& 15& ${0.023}$& ${0.0098}$& $\textit{-0.1400}$& ${0.33}$& ${0.45}$& ${0.68}$& ${0.62}$& ${0.82}$& ${0.96}$\\
\cmidrule(rl){2-2} \cmidrule(rl){3-5} \cmidrule(rl){6-8} \cmidrule(rl){9-11}
& 30& ${0.0032}$& $\textit{-0.043}$& $\textit{-0.0099}$& ${0.25}$& ${0.42}$& ${0.72}$& ${0.56}$& ${0.83}$& ${1.1}$\\
\cmidrule(rl){2-2} \cmidrule(rl){3-5} \cmidrule(rl){6-8} \cmidrule(rl){9-11}
& 90& ${0.013}$& ${0.024}$& $\textit{-0.015}$& ${0.088}$& ${0.21}$& ${0.41}$& ${0.24}$& ${0.48}$& ${0.82}$\\
\cmidrule(rl){2-2} \cmidrule(rl){3-5} \cmidrule(rl){6-8} \cmidrule(rl){9-11}
& 170& ${0.012}$& ${0.016}$& ${0.0085}$& ${0.046}$& ${0.081}$& ${0.11}$& ${0.12}$& ${0.22}$& ${0.29}$\\
\bottomrule
\end{tabular}
}
\caption{Median values of \ac{ORL} for the exploration sets with component masses $(m_1,m_2)=(35,29) M_\odot$. Italic entries denote sets with negative values.}
\label{table:m1_35_m2_29_med_Z}
\end{minipage}
\end{table}